\documentclass[twoside,fleqn,11pt]{article}

\usepackage{fullpage,cite,amsxtra,graphicx}

\numberwithin{equation}{section}
\makeatletter\renewcommand\@biblabel[1]{#1.}\makeatother
\DeclareMathOperator\re{Re}
\DeclareMathOperator\im{Im}
\DeclareMathOperator\sn{sn}
\DeclareMathOperator\dn{dn}
\DeclareMathOperator\cn{cn}
\DeclareMathOperator\sgn{sgn}
\newlength{\tbl}
\newenvironment{tab}{\par\addtolength\tbl{1.2pc}\leftskip\tbl}{\addtolength\tbl{-1.2pc}\par\leftskip\tbl}
\allowdisplaybreaks

\begin{document}

\title{\bf Numerical implementation of the exact dynamics of free rigid bodies}

\author{Ramses van Zon, Jeremy Schofield\\[2mm]
\normalsize\em Chemical Physics Theory Group, Department of Chemistry, University
\\
\normalsize\em of Toronto, 80 St.\ George Street, Toronto, Ontario, M5S 3H6, Canada}

\date{December 15, 2006}

\maketitle

\begin{abstract}
In this paper the exact analytical solution of the motion of a rigid
body with arbitrary mass distribution is derived in the absence of
forces or torques.  The resulting expressions are cast into a form
where the dependence of the motion on initial conditions is explicit
and the equations governing the orientation of the body involve only
real numbers.  Based on these results, an efficient method to
calculate the location and orientation of the rigid body at arbitrary
times is presented.  This implementation can be used to verify the
accuracy of numerical integration schemes for rigid bodies, to
serve as a building block for event-driven discontinuous molecular
dynamics simulations of general rigid bodies, and for
constructing symplectic integrators for rigid body dynamics.  
\end{abstract}

\section{Introduction}

The importance of the dynamics of rigid bodies in physics and
engineering has been recognized since the early 19th century.  The
early work of Euler, Hermite, Poisson, Jacobi and many others on
exactly soluble systems lead to great advances in the field of applied
mathematics as well as in mechanics.  More recently, rigid hard
spheres served as the first prototypical model for fluids treated by
computer simulation\cite{AlderWainwright57}.  Today, rigid bodies are
used to model phenomena on a variety of different length scales: On
molecular length scales, rigid bodies are used in the modeling of the
microscopic dynamics of molecules in condensed
phases\cite{Chapelaetal84,DMD1,DMD2}, on a mesoscopic scale, they are
used to construct simple models of polymers and other complex
systems\cite{FrenkelMaguire83}, while on a macroscopic level they play
an important role in robotics. The dynamics of rigid bodies is also of
relevance to the computer game industry, where many improvements in
simulating rigid bodies have been developed\cite{Animation1}.
Finally, on an even larger scale, many astrophysical objects such as
planets, satellites and space crafts can be regarded as rigid bodies
on certain time scales\cite{Wertz78,Masutanietal94}.

According to the laws of classical mechanics, the motion of a rigid
body consists of translation of the center of mass of the body and
rotation of the orientation of the body about its center of mass.  In
the most general case in which the body is subject to forces and
torques, an analytical solution of the dynamics is not possible and a
numerical scheme is required to integrate the equations of
motion. With the advent of powerful computers, much work has been
devoted to finding stable and efficient integration schemes for
rotating rigid
bodies\cite{rotation0,rotation,rotation2,CelledoniSafstrom}.  The use
of numerical integrators is now so widespread that they are frequently
used even in cases where an analytical solution of the dynamics is
available.  This is unfortunate, since an exact solution can not only
serve as a special case against which numerical integration schemes
may be tested, but can also be directly used in so-called
discontinuous molecular dynamics of rigid bodies, in which the bodies
perform free motion in between interaction events. A hard sphere gas
is the prototypical example of such a system. In these kinds of
simulations, using an exact solution of the free dynamics yields an
enormous computational benefit compared to having to integrate the
free dynamics numerically (see e.g.\ Refs. \citeonline{DMD1} and
\citeonline{DMD2}). Furthermore, for molecular dynamics of systems
with a continuous potential, an exact solution is also often useful in
constructing so-called symplectic integration schemes, and typically
leads to enhanced
stability\cite{appendixC,CelledoniSafstrom,exactpart,newintegrator}.

Perhaps the principal reason why the analytical solution of the motion
of a free rigid body is seldom used is that the general solution is
simply not well-known. Many advanced textbooks in mechanics avoid
discussing the motion of general rigid bodies, and only consider
certain symmetric cases\cite{Goldstein} in which the equations of
motion are particularly simple. This probably is due to the fact that
the general solution --- apparently first found by Rueb\cite{Rueb1834}
in 1834 and later completed by Jacobi in 1849\cite{Jacobi1850} ---
involves special functions, called elliptic and theta functions, which
are perhaps less familiar than other special functions. Furthermore,
even when the case of a general rotor is
discussed\cite{Rueb1834,Jacobi1850,Whittaker,Masutanietal94,MarsdenRatiu},
the treatment of its motion is often incomplete, in rather abstract
form (using complex-valued functions) and presented in a special
inertial coordinate frame rather than a general laboratory inertial
frame.

The goal of this paper is to demonstrate that none of these issues
needs to be an obstacle to the use of the exact solution of the
equations of motion of an asymmetric rigid body in numerical work.  It
will first be shown that although the derivation of the general
solution of the equations of motion of a rigid body requires a bit of
complex analysis, the final result can be expressed without any
reference to complex arithmetic. Secondly, the solution will be
formulated in a general inertial frame and for general initial
conditions of the rigid body. Finally, it will be shown that the
special functions occurring in the solution can be numerically
implemented in an efficient fashion. Based on these considerations,
the numerical implementation of the (admittedly non-trivial) motion of
an asymmetric rigid body in the absence of forces and torques is
relatively straightforward.

The paper is organized as follows: In section~\ref{motion}, the motion
of a rigid body is reviewed starting in subsection \ref{rigid} with a
brief overview of the properties of rigid bodies. Subsequently, in
subsection \ref{dynamics} the equations of motion are given, while in
\ref{force} these are specialized to the case of free motion.  The
novel part of the paper starts in section \ref{explicit}, where a new
derivation of the exact solution of the time dependence of the
orientation of the body is given, first in general and then
explicitly for the spherical top, the symmetric top and the asymmetric
top (sections \ref{sssolution}, \ref{sssolutionb} and
\ref{sssolutionc}, respectively).  In section~\ref{numerical}, some
further numerical issues are addressed.  An example is given in
section~\ref{example} and the paper ends in section~\ref{discussion}
with a discussion of the results and their possible applications.

\section{Review of the motion of rigid bodies}
\label{motion}

\subsection{Rigid bodies}
\label{rigid}

The shape of a rigid body is specified by the set of all material
points $\{\tilde{\boldsymbol{r}}_{i}\}$ of the body. The points
$\tilde{\boldsymbol{r}}_{i}=(\tilde{x}_{i},\tilde{y}_{i},\tilde{z}_{i})$
are three-dimensional vectors with respect to a reference coordinate
frame called the \emph{body frame}, and constitute a reference
orientation of the body\cite{MarsdenRatiu}.  Furthermore, a mass
$m_{i}$ is associated with each material point.  The mass distribution
of the body plays an important role in its dynamics.

Since any translation or rotation acting on all points
$\tilde{\boldsymbol{r}}_{i}$ leaves the shape of the body unchanged,
there is an arbitrariness in the choice of body frame which can be
exploited. By a suitable translation, one may always set the center of
mass to be in the origin, i.e.,
\begin{equation}
\sum_{i} m_{i}\tilde{\boldsymbol{r}}_{i}=0.
\label{cm}
\end{equation}
Furthermore, using a suitable rotation, the moment of inertia tensor
$\tilde{\mathbf{I}}$ can be brought to its principal form:
\begin{equation}
  \tilde{\mathbf{I}} = \sum_{i} m_{i}  \begin{pmatrix}
                   \tilde{y}_{i}^{2}+\tilde{z}_{i}^{2}&-\tilde{x}_{i}\tilde{y}_{i}&-\tilde{x}_{i}\tilde{z}_{i}\\
                   -\tilde{x}_{i}\tilde{y}_{i}&\tilde{x}_{i}^{2}+\tilde{z}_{i}^{2}&-\tilde{y}_{i}\tilde{z}_{i}\\
                   -\tilde{x}_{i}\tilde{z}_{i}&-\tilde{y}_{i}\tilde{z}_{i}&\tilde{x}_{i}^{2}+\tilde{y}_{i}^{2}
	       \end{pmatrix} 
=
  \begin{pmatrix}
    I_{1}&0&0\\
    0&I_{2}&0\\
    0&0&I_{3}
  \end{pmatrix}
.
\label{Iprincipal}
\end{equation}
For subsequent developments, it will be assumed that these
transformations have been performed. 

Because the body frame is fixed to the body, the material points
$\tilde{\boldsymbol{r}}_{i}$ are independent of time. However since
the body itself moves through physical space, the location of each
point $\tilde{\boldsymbol{r}}_{i}$ in the physical coordinate system,
or \emph{lab frame}, must be determined as a function of time to
describe its motion.  The position of mass point $i$ in the lab frame
will be denoted by
$\boldsymbol{r}_{i}(t)=(x_{i}(t),y_{i}(t),z_{i}(t))$. Since the body
is rigid, its motion is a time-dependent
orientation-and-distance-preserving transformation from the body frame
to the lab frame. The most general such transformation is a
combination of a translation and a rotation:
\begin{equation}
  \boldsymbol{r}_{i}(t) = \boldsymbol{R}(t) + \mathbf{A}^{\dagger}(t)\, \tilde{\boldsymbol{r}}_{i}.
\label{position}
\end{equation}
Here and below, matrix-vector and matrix-matrix products such as $
\mathbf{A}^{\dagger}(t)\, \tilde{\boldsymbol{r}}_{i}$ will be denoted
implicitly, i.e, without a ``$\cdot$'', which will only be used for
inner products. For notational simplicity, the explicit time
dependence will be omitted in most expressions, i.e.,
$\boldsymbol{r}_{i}(t)$ will be denoted simply by
$\boldsymbol{r}_{i}$. Exceptions are if the time argument is equal to
zero (e.g.\ $\boldsymbol{r}_{i}(0)$) or integrated over.

In Eq.~\eqref{position}, the vector $\boldsymbol{R}$ denotes the
position of the center of mass at time $t$, while the orthogonal
matrix $\mathbf{A}^{\dagger}$ represents the orientation of the body
with respect to the center of mass at that time and is often called
the \emph{attitude matrix}. Note that $\mathbf{A}$ transforms vectors
from the lab to the body frame, while its transpose
$\mathbf{A}^{\dagger}$ transforms vectors from the body to the lab
frame.

The motions of the different material points of a rigid body are
obviously closely related. Instead of working with the (linear)
velocities of all points, one can instead use a formulation of the
dynamics that utilizes the angular velocity vector
$\boldsymbol{\omega}=(\omega_{1},\omega_{2},\omega_{3})$ of the body
around its center of mass. This angular velocity vector is defined
such that its direction coincides with the rotation axis and its
magnitude coincides with the rate at which it rotates. As a
consequence of this definition, the velocity of any point of the rigid
body satisfies the standard relation \cite{Goldstein}
\begin{equation}
  \boldsymbol{v}_{i} = \boldsymbol{V}+\boldsymbol{\omega}\times(\boldsymbol{r}_{i}-\boldsymbol{R})
\label{velocity0}
\end{equation}
where $\boldsymbol{V}=\dot{\boldsymbol{R}}$ is the velocity of the center of mass.
Defining the antisymmetric matrix
\begin{equation}
   \mathbf{W}(\boldsymbol{\omega}) = \begin{pmatrix}
                        0&-\omega_{3}&\omega_{2}\\
			\omega_{3}&0&-\omega_{1}\\
			-\omega_{2}&\omega_{1}&0
			\end{pmatrix},
\label{skew}
\end{equation}
and using Eq.~\eqref{position}, Eq.~\eqref{velocity0} can also be written as
\begin{equation}
  \boldsymbol{v}_{i} = \boldsymbol{V}+\mathbf{W}(\boldsymbol{\omega})\mathbf{A}^{\dagger}\tilde{\boldsymbol{r}}_{i}.
\label{velocity}
\end{equation}
Taking the time derivative of Eq.~\eqref{position} and using
Eq.~\eqref{velocity}, one sees that $\omega$ and $\dot{\mathbf{A}}$
are related via
\begin{equation}
  \dot{\mathbf{A}}^{\dagger}\,  \mathbf{A} = \mathbf{W}(\boldsymbol{\omega}),
\label{W}
\end{equation}

Given the central role of rotation matrices below, it is useful to
establish some notation.  A rotation matrix $\mathbf{U}$ is a special
orthogonal matrix that can be specified by a rotation axis
$\hat{\boldsymbol n}=(n_{1},n_{2},n_{3})$ and an angle $\psi$. Here
$\hat{\boldsymbol n}$ is a unit vector, so that one may also say that
any non-unit vector $\psi\hat{\boldsymbol n}$ can be used to specify a
rotation, where its norm is equal to the angle $\psi$ and its
direction is along the axis $\hat{\boldsymbol n}$. In fact, one can
express this rotation matrix as $\mathbf{U}(\psi \hat{\boldsymbol
n})=\exp\,\mathbf{W}(\psi\hat{\boldsymbol n})$.  The explicit form of
this rotation matrix may be found using Rodrigues'
formula\cite{Goldstein}.

\subsection{Dynamics}
\label{dynamics}

It is clear from Eq.~\eqref{position} that the motion of a rigid body
as a function of time is determined by the time dependence of the
center of mass vector $\boldsymbol{R}$ and the attitude matrix
$\mathbf{A}$. According to the mechanics of rigid bodies, these follow
from the equations of motion
\begin{align}
\boldsymbol{F}&=  \dot{\boldsymbol{P}} 
&
\boldsymbol{\tau} &= \dot{\boldsymbol{L}}.
\end{align}
Here, $\boldsymbol{F}$ is the sum of all forces acting on the body,
$\boldsymbol{P}=M\boldsymbol{V}$ (with $M=\sum_{i} m_{i}$) is the
total momentum, $\boldsymbol{\tau}$ the total torque with respect to
the center of mass of the body and $\boldsymbol{L}=\mathbf{I}\,
\boldsymbol{\omega}$ is the angular momentum. One may equivalently
write
\begin{align}
 \boldsymbol{F} &= M \dot{\boldsymbol{V}}
&
\boldsymbol{\tau} &=  \frac{d}{dt}(\mathbf{I}\, \boldsymbol{\omega}).
\end{align}
The latter equation is more conveniently written in the body frame
using $\tilde{\boldsymbol{\omega}}=\mathbf{A}\, \boldsymbol{\omega}$,
$\tilde{\boldsymbol{\tau}}=\mathbf{A}\, \boldsymbol{\tau}$ and
$\tilde{\mathbf{I}}=\mathbf{A}\, \mathbf{I}\, \mathbf{A}^{\dagger}$,
which yields
\begin{align}
  \tilde{\boldsymbol{\tau}} &=
  \mathbf{A}\, \dot{\mathbf{A}}^{\dagger}\, \tilde{\mathbf{I}}\, \tilde{\boldsymbol{\omega}}
  + \tilde{\mathbf{I}}\, \dot{\tilde{\boldsymbol{\omega}}}
\nonumber\\*
&
=  \mathbf{A}\, \mathbf{W}(\boldsymbol{\omega})\,
  \mathbf{A}^{\dagger}\, 
  \tilde{\mathbf{I}}\, \tilde{\boldsymbol{\omega}}
+ \tilde{\mathbf{I}}\, \dot{\tilde{\boldsymbol{\omega}}}
\nonumber\\*
&
=
  \mathbf{W}(\tilde{\boldsymbol{\omega}})\, \tilde{\mathbf{I}}\, \tilde{\boldsymbol{\omega}}
+ \tilde{\mathbf{I}}\, \dot{\tilde{\boldsymbol{\omega}}}
,
\label{almostEuler}
\end{align}
where Eq.~\eqref{W} was used to obtain the second equality, and the
third equality was obtained using
\begin{equation}
 \mathbf{A}\, \mathbf{W}(\boldsymbol{\omega})\,  \mathbf{A}^{\dagger} 
= \mathbf{W}(\mathbf{A}\, \boldsymbol{\omega})
= \mathbf{W}(\tilde{\boldsymbol{\omega}}). 
\label{relation}
\end{equation}
Writing out Eq.~\eqref{almostEuler} in its components gives the
so-called \emph{Euler equations}.  Solving these equations yields the
time dependence of the angular velocities in the body frame. To
consequently find the attitude matrix $\mathbf{A}$, one uses
Eqs.~\eqref{W} and \eqref{relation} to find
\begin{equation}
  \dot{\mathbf{A}} = - \mathbf{W}(\tilde{\boldsymbol{\omega}})\,  \mathbf{A}.
\label{EOM}
\end{equation}

\subsection{Force and torque-free case}
\label{force}

In the special case where all external forces and torques are zero,
i.e., $\boldsymbol{F} = 0$ and $\boldsymbol{\tau} =0$, the equations
of motion to solve simplify to
\begin{align}
  \dot{\boldsymbol{V}} &= 0
&
\tilde{\mathbf{I}}\, \dot{\tilde{\boldsymbol{\omega}}} &= - 
  \tilde{\boldsymbol{\omega}}\times \tilde{\mathbf{I}}\, \tilde{\boldsymbol{\omega}}
&
  \dot{\mathbf{A}} = - \mathbf{W}(\tilde{\boldsymbol{\omega}})\,  \mathbf{A}.  
\label{threequations}
\end{align}
In the absence of forces and torques, the dynamics of the system is
invariant under rotations and under translations in time and space.
As a consequence of these symmetries, the energy $E$, momentum
$\boldsymbol{P}$ and angular momentum $\boldsymbol{L}$ are conserved,
where $E$ is given by
\begin{equation}
  E = E_{T} + E_{R}
\end{equation}
with the translational and rotational energies equal to
\begin{equation}
\begin{split}
  E_{T} &= \frac{1}{2M}|\boldsymbol{P}|^{2} = \frac{1}{2}M|\boldsymbol{V}|^{2}
\\
  E_{R} &=
  \frac12\boldsymbol{L}\cdot\mathbf{I}^{-1}\, \boldsymbol{L}
  =
  \frac12\boldsymbol{\omega}\cdot\mathbf{I}\,  \boldsymbol{\omega} 
  = 
  \frac12\tilde{\boldsymbol{\omega}}\cdot\tilde{\mathbf{I}}\, \tilde{\boldsymbol{\omega}}
  =
  \frac12(I_{1}\tilde{\omega}^{2}_{1}+I_{2}\tilde{\omega}^{2}_{2}+I_{3}\tilde{\omega}^{2}_{3}),
\label{ERexpr}
\end{split}
\end{equation}
respectively.

The time dependence of the translational part of the motion follows
from the equation of motion $\dot{\boldsymbol{V}}=0$, which is easily
solved to obtain
\begin{equation}
\begin{split}
   \boldsymbol{V} &= \boldsymbol{V}(0)
\\*
   \boldsymbol{R} &= \boldsymbol{R}(0) + \boldsymbol{V}(0) t.
\label{Rt}
\end{split}
\end{equation}
Since $\boldsymbol{V}(t) = \boldsymbol{V}(0)$, the translational
energy $E_{T}$ is conserved in the dynamics, which, in turn, implies
that the rotational energy $E_{R}$ is also conserved.

To solve the rotational equations of motion is less trivial.  One has
to solve the middle equation of \eqref{threequations}, which, written
out in components, reads
\begin{equation}
\begin{split}
  I_{1}\dot{\tilde{\omega}}_{1} &= \tilde{\omega}_{2}\tilde{\omega}_{3}(I_{2}-I_{3})
\\
  I_{2}\dot{\tilde{\omega}}_{2} &= \tilde{\omega}_{3}\tilde{\omega}_{1}(I_{3}-I_{1})
\\
  I_{3}\dot{\tilde{\omega}}_{3} &= \tilde{\omega}_{1}\tilde{\omega}_{2}(I_{1}-I_{2}).
\end{split}
\label{EE}
\end{equation}
The general solution of this set of equations will be presented below.
Given this solution, the remaining task is to determine the attitude
matrix $\mathbf{A}$ from Eq.~\eqref{EOM}.

Although the general solution for $\mathbf{A}$ is hard to find in any
textbook, it can be found in Jacobi's treatise on rigid body motion
from 1849\cite{Jacobi1850}. Unfortunately, for mathematical elegance,
the attitude matrix was only specified up to a rotation at a uniform
speed which was not clearly identified, so that its numerical
implementation is not altogether clear. Furthermore, Jacobi's
derivation relies heavily on geometric arguments and Euler angles,
which often pose problems in numerical applications. For this reason,
the next section contains a novel derivation of the exact solution of
$\mathbf{A}$ without reference to Euler angles, leading to an
expression which is more readily implemented.

\section{Exact solution of the attitude matrix in the absence of torques}
\label{explicit}

While discussions of how to solve the Euler equations in the body
frame are common, one rarely sees any mention of how to go about
integrating the equation for the attitude matrix. Therefore, we have
chosen to treat this derivation in a bit more detail than one would
perhaps expect from a computational paper.  First the general form of
the solution will be derived, after which the three different case of
rigid bodies are explicitly considered: spherical bodies (section
\ref{sssolution}), symmetric tops (section \ref{sssolutionb}), and
asymmetric tops (section \ref{sssolutionc}).

To obtain the general form of the attitude matrix $\mathbf{A}$, we
have to solve Eq.~\eqref{EOM}.  Since this is a linear equation, its
solution can be written in the form
\begin{equation}
  \mathbf{A} = \mathbf{P}\, \mathbf{A}(0).
\label{genAmat}
\end{equation}
Here, $\mathbf{P}$ is a time dependent matrix which also satisfies
Eq.~\eqref{EOM}, but with initial condition
$\mathbf{P}(0)=\mathbf{1}$.  This corresponds to a case in which the
body and lab frame initially coincide, i.e., in which the body is in
the ``upright'' position.  In this upright lab frame, the angular
momentum vector is given by
\begin{equation}
  \boldsymbol{L}
  =\begin{pmatrix}L_{1}\\L_{2}\\L_{3}\end{pmatrix}
  =\tilde{\boldsymbol{L}}(0) .
\end{equation}

It will prove to be more convenient to work in a different lab frame,
called the {\it invariant frame} (with vectors in this frame denoted
by primed quantities), in which the angular momentum vector is along
the $z$-axis and is equal to $\boldsymbol{L}'=(0,0,L)^{\dagger}$. Such
a frame can be found by performing a rotation of the original frame
through a rotation matrix $\mathbf{T}^{\prime\dagger}_{1}(0)$ (the
reason for the peculiar notation will become clear below).  This
rotation is not unique and can be chosen such that one first rotates
around the $z$-axis until the rotated angular momentum vector no
longer has a $y$-component, and then subsequently rotates around the
$y$-axis to remove the $x$-component of the angular momentum vector as
well. Denoting $\tilde{L}_{\perp} =
[{\tilde{L}_{1}}^{2}+{\tilde{L}_{2}}^{2}]^{1/2}$, the combined
rotation is easily shown to be\footnote{For the special case when
$\tilde{L}_{\perp}(0)=0$ one may take $\mathbf{T}'_{1}(0) = {\rm
diag}(\pm1,1,\pm1)$ where the sign is chosen according to
${\tilde{L}_{3}(0)}/{L}=\pm1$, i.e., depending on whether in the
original lab frame $\boldsymbol{L}$ pointed in the positive or the
negative $z$-direction.}
\begin{equation}
  \mathbf{T}^{\prime\dagger}_{1}(0) 
  = 
\underbrace{
  \begin{pmatrix}
  \frac{\tilde{L}_{3}(0)}L  &  0  & -\frac{\tilde{L}_{\perp}(0)}{L}\\
  0                         &  1  & 0\\
  \frac{\tilde{L}_{\perp}(0)}{L}  & 0  &  \frac{\tilde{L}_{3}(0)}{L}
  \end{pmatrix}}_{\text{rotation to get $L'_{x}(0)=0$}}
  \, 
\underbrace{
  \begin{pmatrix}
  \frac{\tilde{L}_{1}(0)}{\tilde{L}_{\perp}(0)} & \frac{\tilde{L}_{2}(0)}{\tilde{L}_{\perp}(0)} &0\\
  -\frac{\tilde{L}_{2}(0)}{\tilde{L}_{\perp}(0)} & \frac{\tilde{L}_{1}(0)}{\tilde{L}_{\perp}(0)} &0\\
  0             &          0           &         1
  \end{pmatrix}
}_{\text{rotation to get $L'_{y}(0)=0$}}
=
\begin{pmatrix}
  \frac{\tilde{L}_{1}(0)\tilde{L}_{3}(0)}{\tilde{L}_{\perp}(0) L}&
  \frac{\tilde{L}_{2}(0)\tilde{L}_{3}(0)}{L\tilde{L}_{\perp}(0)} &
  -\frac{\tilde{L}_{\perp}(0)}{L}\\
  -\frac{\tilde{L}_{2}(0)}{\tilde{L}_{\perp}(0)}  & 
  \frac{\tilde{L}_{1}(0)}{\tilde{L}_{\perp}(0)} & 0\\
  \frac{\tilde{L}_{1}(0)}{L}&
  \frac{\tilde{L}_{2}(0)}{L}&
  \frac{\tilde{L}_{3}(0)}{L}
\end{pmatrix}
.\label{R10}
\end{equation}
This rotation is defined such that if $\boldsymbol{v}$ is a vector in
the original lab frame, then
$\boldsymbol{v}'=\mathbf{T}^{\prime\dagger}_{1}(0)\, \boldsymbol{v}$
is the corresponding vector in the invariant frame.  To find the
matrix corresponding to $\mathbf{P}$ in this new frame, note that
since $\mathbf{P}$ relates vectors to the body frame through
$\tilde{\boldsymbol{v}}=\mathbf{P}\, \boldsymbol{v}$, so that
$\tilde{\boldsymbol{v}}=\mathbf{P}\, \mathbf{T}'_{1}(0)\,
\boldsymbol{v}'=\mathbf{P}'\, \boldsymbol{v}'$, one can identify
\begin{equation}
  \mathbf{P}'=\mathbf{P}\,  \mathbf{T}'_{1}(0). 
\end{equation}
Since $\mathbf{T}'_{1}(0)$ is a constant matrix multiplying on the
right, $\mathbf{P}'$ also satisfies Eq.~\eqref{EOM} with initial
condition $\mathbf{P}'(0)=\mathbf{T}'_{1}(0)$.

The convenience of this choice of frame becomes clear when
$\mathbf{P}'$ is written in terms of its columns,
\begin{equation}
  \mathbf{P}'=\Big[\hat{\boldsymbol u}_{1}\:
  \hat{\boldsymbol u}_{2}\:\hat{\boldsymbol u}_{3}\Big]                     
\label{columns}
\end{equation}
and one notes that 
\begin{equation}
  \hat{\boldsymbol u}_{3}=\mathbf{P}'\, 
\left(\begin{smallmatrix}0\\0\\1\end{smallmatrix}\right)
  =\frac{\mathbf{P}'\, \boldsymbol{L}'}{L}= \frac{\tilde{\boldsymbol{L}}}{L}
,
\end{equation}
the components of which are known once the Euler equations are solved.
Thus, by this choice of frame, one is able to determine the third
column of the matrix $\mathbf{P}'$.

The remaining elements of $\mathbf{P}'$ can all be expressed in terms
of a single time-dependent angle $\psi$. This is due to the
orthogonality of $\mathbf{P}'$, which implies that the other two
columns $\hat{\boldsymbol u}_{1}$ and $\hat{\boldsymbol u}_{2}$ must
lie in a plane orthogonal to $\hat{\boldsymbol u}_{3}$ and must also
be orthogonal to each other. Denoting $\hat{\boldsymbol e}_{1}$ and
$\hat{\boldsymbol e}_{2}$ as two chosen orthogonal unit vectors in
this plane, one can therefore write
\begin{align}
  \hat{\boldsymbol u}_{1} &= \hat{\boldsymbol e}_{1}\cos\psi 
  -\hat{\boldsymbol e}_{2}\sin\psi
  \label{P1}
  \\
  \hat{\boldsymbol u}_{2} &= \hat{\boldsymbol e}_{1}\sin\psi
 +\hat{\boldsymbol e}_{2}\cos\psi
  \label{P2},
\end{align}
where the unit-vectors $\hat{\boldsymbol e}_{1}$ and $\hat{\boldsymbol
e}_{2}$ are chosen to be:\footnote{For the special case when
$\tilde{L}_{\perp}=0$ one may take $\hat{\boldsymbol
e}_{2}=(0,1,0)^{\dagger}$ and $\hat{\boldsymbol e}_{1} =
\hat{\boldsymbol e}_{2}\times \hat{\boldsymbol u}_{3}$.}
\begin{align}
  \hat{\boldsymbol e}_{2} &= \frac{\hat{\boldsymbol e}_{z}\times \hat{\boldsymbol u}_{3}}
	           {|\hat{\boldsymbol e}_{z}\times \hat{\boldsymbol u}_{3}|}
  = \frac{\hat{\boldsymbol e}_{z}\times \tilde{\boldsymbol{L}}}
  {|\hat{\boldsymbol e}_{z}\times \tilde{\boldsymbol{L}}|}
  =\begin{pmatrix}
  -\frac{\tilde{L}_{2}}{\tilde{L}_{\perp}}\\
  \frac{\tilde{L}_{1}}{\tilde{L}_{\perp}}\\
  0                                          
  \end{pmatrix}
 \label{e1}
\\
  \hat{\boldsymbol e}_{1} &= \hat{\boldsymbol e}_{2}\times \hat{\boldsymbol u}_{3}
  =
  \begin{pmatrix}
  \frac{\tilde{L}_{1}\tilde{L}_{3}}{L\tilde{L}_{\perp}}\\
  \frac{\tilde{L}_{2}\tilde{L}_{3}}{L\tilde{L}_{\perp}}\\
  -\frac{\tilde{L}_{\perp}}{L},
  \end{pmatrix}
                                                            \label{e2}
\end{align}
where $\hat{\boldsymbol e}_{z}=(0,0,1)$, and we have used the fact
that $|\hat{\boldsymbol e}_{z}\times
\tilde{\boldsymbol{L}}|=(\tilde{L}_{1}^{2}+
\tilde{L}_{2}^{2})^{1/2}=\tilde{L}_{\perp}$.  Other choices of
orthogonal unit-vectors $\hat{\boldsymbol e}_{1}$ and
$\hat{\boldsymbol e}_{2}$ change the as-of-yet undetermined time
dependent angle $\psi$ by a time independent offset. The current
choice has the advantage that the matrix $[\hat{\boldsymbol
e}_{1}(0)\:\hat{\boldsymbol e}_{2}(0)\:\hat{\boldsymbol u}_{3}(0)]$ is
identical to $\mathbf{T}'_{1}(0)$, so that $\mathbf{P}'(0) =
\mathbf{T}'_{1}(0)$, and hence
\begin{equation}
  \psi(0) = 0.                                            \label{psi0}
\end{equation}
Using Eqs.~\eqref{P1} and \eqref{P2}, $\mathbf{P}'$ has effectively
been written as a product of two rotation matrices,
\begin{equation}
  \mathbf{P}' = \mathbf{T}_{1}' \,  \mathbf{T}_{2}',
\end{equation}
where
\begin{align}
  \mathbf{T}_{1}' &=
  \Big[ \hat{\boldsymbol e}_{1}\: \hat{\boldsymbol e}_{2}\: \hat{\boldsymbol u}_{3}\Big]
=
  \begin{pmatrix}
  \frac{\tilde{L}_{1}\tilde{L}_{3}}{L\tilde{L}_{\perp}} &
  -\frac{\tilde{L}_{2}}{\tilde{L}_{\perp}} & 
  \frac{\tilde{L}_{1}}{L}\\
  \frac{\tilde{L}_{2}\tilde{L}_{3}}{L\tilde{L}_{\perp}} &
  \frac{\tilde{L}_{1}}{\tilde{L}_{\perp}} & 
  \frac{\tilde{L}_{2}}{L}\\
  -\frac{\tilde{L}_{\perp}}{L}  &
  0  &
  {\frac{\tilde{L}_{3}}{L}}
  \end{pmatrix}
  \label{R1}
\\
  \mathbf{T}'_{2} &=
  \begin{pmatrix}
     \cos\psi&\sin\psi&0\\
     -\sin\psi&\cos\psi&0\\
     0&0&1 
  \end{pmatrix}
     =\mathbf{U}(-\psi\hat{\boldsymbol z}).
     \label{R2}
\end{align}
Note that $\mathbf{T}_{1}'$ in Eq.~\eqref{R1} is a rotation matrix
because $\hat{\boldsymbol e}_{1}$, $\hat{\boldsymbol e}_{2}$ and
$\hat{\boldsymbol u}_{3}$ form an orthogonal set by their construction
in Eqs.~\eqref{e1} and \eqref{e2}.  The matrix $\mathbf{P}$ can
therefore be written as
\begin{equation}
  \mathbf{P} =
  \mathbf{T}'_{1}\,  \mathbf{T}'_{2}\,  \mathbf{T}_{1}^{\prime\dagger}(0).
\label{Pinprime}
\end{equation}
For the special cases of the spherical top and symmetric top, it is
more convenient to express $\mathbf{P}$ as a product of two rotation
matrices (implicitly also found in Ref.~\citeonline{Masutanietal94}):
\begin{equation}
  \mathbf{P} = \mathbf{T}_{1} \,  \mathbf{T}_{2},
  \label{Pmat}
\end{equation}
where
\begin{align}
  \mathbf{T}_{1}&=  \mathbf{T}'_{1}\, \mathbf{T}^{\prime\dagger}_{1}(0)
\label{Q1}
\\
  \mathbf{T}_{2} &= \mathbf{T}'_{1}(0)\, \mathbf{T}'_{2}\, 
                           \mathbf{T}^{\prime\dagger}_{1}(0).
\label{Q2orig}
\end{align}
Note that $\mathbf{T}_{1}$ and $\mathbf{T}'_{1}$ are determined by the
solution of the Euler equations and require no further manipulations
once the general solution of $\tilde{\boldsymbol{L}}$ is known.
 
Turning next to the matrix $\mathbf{T}_{2}$, noting that for any
rotation $\mathbf{R}$ and vector $\boldsymbol{x}$, $\mathbf{R}\,
\mathbf{U}(\boldsymbol{x})\, \mathbf{R}^{\dagger} =
\mathbf{U}(\mathbf{R}\, \boldsymbol{x})$ and that
$\mathbf{T}'_{1}(0)\, \hat{\boldsymbol
z}=\tilde{\boldsymbol{L}}(0)/L$, the matrix $\mathbf{T}_{2}$ can
written as a rotation by an angle of $-\psi$ around the axis
$\tilde{\boldsymbol{L}}(0)/L$, i.e., using Eqs.~\eqref{R2} and
\eqref{Q2orig},
\begin{equation}
\mathbf{T}_{2} = \mathbf{U}(-\psi\tilde{\boldsymbol{L}}(0)/L).
\label{Q2}
\end{equation}

The final task to determine $\mathbf{P}$ consists of deriving a
differential equation for the time dependent angle $\psi$ and solving
it subject to the initial condition $\psi(0)=0$
(cf.~Eq.~\eqref{psi0}). To obtain this differential equation, note
that from Eqs.~\eqref{EOM}, \eqref{W} and \eqref{columns}, it follows
that
\begin{equation*}
  \dot{\hat{\boldsymbol{u}}}_{1} = 
  -\tilde{\boldsymbol{\omega}}\times\hat{\boldsymbol u}_{1},
\end{equation*}
and hence from Eq.~\eqref{P1} one finds
\begin{equation*}
  \dot{\hat{\boldsymbol{e}}}_{1}\cos\psi-\hat{\boldsymbol e}_{1}\dot{\psi}\sin\psi
  -\dot{\hat{\boldsymbol{e}}}_{2}\sin\psi-\hat{\boldsymbol e}_{2}\dot{\psi}\cos\psi
= -\tilde{\boldsymbol{\omega}}\times\hat{\boldsymbol e}_{1}\cos\psi
      +\tilde{\boldsymbol{\omega}}\times\hat{\boldsymbol e}_{2}\sin\psi.
\end{equation*}
Taking the inner product with $\hat{\boldsymbol e}_{2}$, using that
$\hat{\boldsymbol
e}_{2}\cdot\dot{\hat{\boldsymbol{e}}}_{2}=(1/2)d|\hat{\boldsymbol
e}_{2}|^{2}/dt=0$ and dividing by $\cos\psi$\footnote{When
$\cos\psi=0$, one can instead take the inner product with
$\hat{\boldsymbol e}_{1}$ and divide by $\sin\psi$, with the same
result.}, yields the differential equation for the angle, $\dot{\psi}
= \Omega$, where the time dependent frequency $\Omega$ can be
expressed as
\begin{align*} 
  \Omega &=  
-\hat{\boldsymbol e}_{1}\, \dot{\hat{\boldsymbol{e}}}_{2}
             +\hat{\boldsymbol e}_{2}\cdot(\tilde{\boldsymbol{\omega}}\times\hat{\boldsymbol e}_{1})
=  -\hat{\boldsymbol e}_{1}\cdot\dot{\hat{\boldsymbol{e}}}_{2}
	     +\tilde{\boldsymbol{\omega}}\cdot(\hat{\boldsymbol e}_{1}\times\hat{\boldsymbol e}_{2})
\:=  -\hat{\boldsymbol e}_{1}\cdot\dot{\hat{\boldsymbol{e}}}_{2}
             +\tilde{\boldsymbol{\omega}}\cdot\hat{\boldsymbol u}_{3}
\nonumber\\*&=  -\hat{\boldsymbol e}_{1}\cdot\dot{\hat{\boldsymbol{e}}}_{2}
             +\tilde{\boldsymbol{\omega}}\cdot\tilde{\boldsymbol{L}}/L.
\end{align*}
It follows from Eq.~\eqref{e1} and \eqref{e2} that
\begin{equation*}
  \hat{\boldsymbol e}_{1}\cdot\dot{\hat{\boldsymbol{e}}}_{2} 
  = \frac{\tilde{L}_{3}(\tilde{L}_{2}\dot{\tilde{L}}_{1}-\tilde{L}_{1}\dot{\tilde{L}}_{2})}{L\tilde{L}_{\perp}^{2}}
.
\end{equation*}
Using Eq.~\eqref{EE} yields $\dot{\tilde{L}}_{1} = \tilde{\omega}_{3}\tilde{L}_{2} -
\tilde{\omega}_{2}\tilde{L}_{3}$ and $\dot{\tilde{L}}_{2} =
\tilde{\omega}_{1}\tilde{L}_{3} - \tilde{\omega}_{3}\tilde{L}_{1}$, whence $\Omega
= L(\tilde{L}_{1}\tilde{\omega}_{1} + \tilde{L}_{2}\tilde{\omega}_{2}) /
\tilde{L}_{\perp}^{2}$, or, expressed in terms of conserved quantities
and $\tilde{\omega}_{3}$ only,
\begin{align}
 \Omega
&  = \frac{L(2E_{R}-I_{3}{\tilde{\omega}_{3}}^{2})}{L^{2}-I_{3}^{2}{\tilde{\omega}_{3}}^{2}}.
           \label{Omegae} 
\end{align}
The angle $\psi$ is then given by
\begin{equation}
  \psi = \int_{0}^{t}\!dt'\:\Omega(t').
\label{psi}
\end{equation}

The formal result for the matrix $\mathbf{P}$, given by
Eqs.~\eqref{R1}--\eqref{Q2}, \eqref{Omegae} and \eqref{psi}, still
contains a time integral for $\psi$ whose integrand depends on the
angular velocity component $\tilde{\omega}_{3}$.  Hence the time
dependence of this component of the angular velocity must be specified
to perform the integral and obtain $\psi$.  The solution of the
components of the angular velocity in the body frame follows from the
Euler equations, which can be analyzed in three separate cases
depending on the values of the principal moments of inertia.

\subsection{Spherical top}
\label{sssolution}

For a spherical top, all moments of inertia are the same:
$I_{1}=I_{2}=I_{3}$, and the Euler equations in Eq.~\eqref{EE} become
extremely simple, namely, $\dot{\tilde{\omega}}_{j} = 0$.  For the
spherical top system, all components of the angular velocity in the
body frame are therefore constant, and hence the components of the
angular momentum in the body frame are constant as well. It therefore
follows that the matrix $\mathbf{T}'_{1}$ in Eq.~\eqref{R1} remains
constant in time, so that
\begin{equation}
\mathbf{T}_{1}=\mathbf{T}'_{1}\, \mathbf{T}_{1}^{\prime\dagger}(0)=\mathbf{1}. 
\end{equation}
The frequency $\Omega$ can also easily be determined by noting that
$\tilde{\omega}_{3}$ is constant and $I_{1}=I_{2}=I_{3}$, so that Eq.~\eqref{Omegae} gives
\begin{equation}
  \Omega 
  =  \frac{L(I_{1}\tilde{\omega}_{1}^{2}+I_{2}\tilde{\omega}_{2}^{2})}
          {I_{1}^{2}\tilde{\omega}_{1}^{2}+I_{2}^{2}\tilde{\omega}_{2}^{2}}
  = \frac{L}{I_{1}} = |\boldsymbol{\omega}|,
\end{equation}
leading to $\psi=\Omega t=|\omega|t$. Equation \eqref{Q2} then gives, with
$\boldsymbol{L}(0)/L=\boldsymbol{\omega}/\Omega$,
\begin{equation}
\mathbf{T}_{2} 
  =
\mathbf{U}(-\boldsymbol{\omega} t),
\end{equation}
so that  one recovers the well-known result that for a spherical rotor
\begin{equation}
\mathbf{P}= \mathbf{T}_{1} \, \mathbf{T}_{2} 
  =
\mathbf{U}(-\boldsymbol{\omega} t).
\label{Pmatrix}
\end{equation}
The rotation matrix $\mathbf{P}$ corresponds to a rotation by an angle
of $-|\boldsymbol{\omega}| t$ around the axis
$\boldsymbol{\omega}/|\boldsymbol{\omega}|$.  Note that the minus sign
in the angle arises here from the fact that if the body rotates one
way, the lab frame, as seen from the body frame, rotates in the
opposite way.

\subsection{Symmetric top}
\label{sssolutionb}

For the case of a symmetric top, $I_{1}=I_{2}$ but $I_{2}\neq
I_{3}$. In that case, one can solve the Euler equations \eqref{EE} in
terms of well-known functions:
\begin{equation}
\begin{split}
  \tilde{\omega}_{1} &= \tilde{\omega}_{1}(0)\cos \omega_{p}t 
  +\tilde{\omega}_{2}(0)\sin\omega_{p} t                        
\\
  \tilde{\omega}_{2} &=  -\tilde{\omega}_{1}(0)\sin \omega_{p}t 
  +\tilde{\omega}_{2}(0)\cos\omega_{p} t                        
\\
  \tilde{\omega}_{3} &= \tilde{\omega}_{3}(0).                 \label{symo}
\end{split}
\end{equation}
Here, the precession frequency $\omega_{p}$ is given by $\omega_{p} =
\left(1-I_{3}/{I_{1}}\right)\tilde{\omega}_{3}(0)$.  From these
equations, it is evident that
$\tilde{L}_{\perp}=I_{1}[\tilde{\omega}_{1}^{2}+\tilde{\omega}_{2}^{2}]^{1/2}$
is conserved for a symmetric top, which allows one to rewrite
$\mathbf{T}_{1}$ in Eq.~\eqref{Q1} as a rotation around the $z$-axis
by an angle of $-\omega_{p} t$:
\begin{equation}
  \mathbf{T}_{1}
=
  \begin{pmatrix}
  \cos\omega_{p} t &\sin\omega_{p} t &0\\
  -\sin\omega_{p} t&\cos\omega_{p} t&0\\
  0&0&1
  \end{pmatrix}=\mathbf{U}(-\omega_{p} t\hat{\boldsymbol z}).
\end{equation}
The final rotation angle $\psi$ can be determined by noting that
\begin{equation}
  \Omega 
  =  \frac{L(I_{1}\tilde{\omega}_{1}^{2}+I_{2}\tilde{\omega}_{2}^{2})}
          {I_{1}^{2}\tilde{\omega}_{1}^{2}+I_{2}^{2}\tilde{\omega}_{2}^{2}}
  = \frac{L}{I_{1}}.
\end{equation}
Again this is a constant but, in contrast to the spherical case, no
longer equal to $|\boldsymbol{\omega}|$, so
\begin{equation}
  \psi = \Omega t.
\end{equation}
The second rotation $\mathbf{T}_{2}$ in Eq.~\eqref{Pmat} is therefore given by
Eq.~\eqref{Q2} with $\psi= L t/I_{1}$, i.e.,
\begin{equation}
  \mathbf{T}_{2} = \mathbf{U}(-\tilde{\boldsymbol{L}}(0)t/I_{1}).
\end{equation}
As a result, the rotation matrix $\mathbf{P}$ for a symmetric top is
given by
\begin{equation}
  \mathbf{P} = \mathbf{T}_{1}\, \mathbf{T}_{2} =  
\mathbf{U}(-\omega_{p} t\hat{\boldsymbol z})\, \mathbf{U}(-\tilde{\boldsymbol{L}}(0)t/I_{1})
\label{Pmatrixsym}
.
\end{equation}

\subsection{Asymmetric top}
\label{sssolutionc}

For a general asymmetric top, all moments of inertia are distinct:
$I_{1}\neq I_{2}\neq I_{3}\neq I_{1}$. The moments of inertia can be
ordered in increasing order of magnitude, and we will choose to call
the middle one always $I_{2}$, while either $I_{1}$ or $I_{3}$ is the
largest one.  In the absence of forces and torques, Eq.~\eqref{EE} is
integrable since the energy $E_{R}$ and the norm of the angular
momentum $L$ are conserved quantities in the body frame.  To ensure
all quantities occurring in the solutions are real-valued (rather than
complex-valued), one needs to consider the quantities $E_{R}$ and
$L^{2}/(2I_{2})$ and make sure that\cite{Jacobi1850}
\begin{equation}
\label{sure} 
\begin{split}
 I_{1}>I_{2}>I_{3}&\qquad\text{\ \ if\ \ } E_{R}>\frac{L^{2}}{2I_{2}}\\
  I_{1}<I_{2}<I_{3}&\qquad\text{\ \ if\ \ } E_{R}<\frac{L^{2}}{2I_{2}}.
\end{split}
\end{equation}
We will refer to this as the Jacobi ordering.  Either situation can
always be realized by choosing which principal axis of the body to
call the first, second or third.

Once the Jacobi ordering is adopted, the solution of the Euler
equations is given by\cite{Jacobi1850,Whittaker,MarsdenRatiu}
\begin{align}
  \tilde{\omega}_{1}
 &= \tilde{\omega}_{1m} \cn\big( \omega_{p}t+\varepsilon|m\big)
\label{omega1t}
\\
  \tilde{\omega}_{2}
 &= \tilde{\omega}_{2m} \sn\big( \omega_{p}t+\varepsilon|m\big)
\label{omega2t}
\\
  \tilde{\omega}_{3}
 &= \tilde{\omega}_{3m} \dn\big( \omega_{p}t+\varepsilon|m\big),
\label{omega3t}
\end{align}
where $\sn$, $\cn$ and $\dn$ are Jacobi elliptic
functions\cite{AbramowitzStegun,WhittakerWatson,Knopp2}, and
\begin{align}
  \tilde{\omega}_{1m} &= \sgn(\tilde{\omega}_{1}(0))\sqrt{\frac{L^{2}-2I_{3}E_{R}}{I_{1}(I_{1}-I_{3})}}
\label{sign1}
\\
  \tilde{\omega}_{2m} &= -\sgn(\tilde{\omega}_{1}(0))\sqrt{\frac{L^{2}-2I_{3}E_{R}}{I_{2}(I_{2}-I_{3})}}
\label{sign2}
\\
  \tilde{\omega}_{3m} &=
  \sgn(\tilde{\omega}_{3}(0))\sqrt{\frac{L^{2}-2I_{1}E_{R}}{I_{3}(I_{3}-I_{1})}}
\label{sign3}
\\
  \omega_{p} &= 
  \sgn(I_{2}-I_{3})  \sgn(\tilde{\omega}_{3}(0))
\sqrt{\frac{(L^{2}-2I_{1}E_{R})(I_{3}-I_{2})}{I_{1}I_{2}I_{3}}}.
\label{omp}
\\
  m &= \frac{(L^{2}-2I_{3}E_{R})(I_{1}-I_{2})}{(L^{2}-2I_{1}E_{R})(I_{3}-I_{2})},
\label{modulus}
\\
  \varepsilon &= -\omega_{p} t_{0} = F(\tilde{\omega}_{2}(0)/\tilde{\omega}_{2m}|m),
\label{epsdef}
\end{align}
where in the last equation, $F$ is the incomplete elliptic integral of
the first kind\cite{AbramowitzStegun}, defined as\footnote{We deviate
here from the somewhat more usual notation $F(\alpha|m)=F(x|m)$ where
$x=\sin\alpha$.}
\begin{align}
F(x|m) = \int_{0}^{x}\frac{dt}{\sqrt{1-m t^{2}}\sqrt{1-t^{2}}}.
\end{align}

\begin{figure}[h]
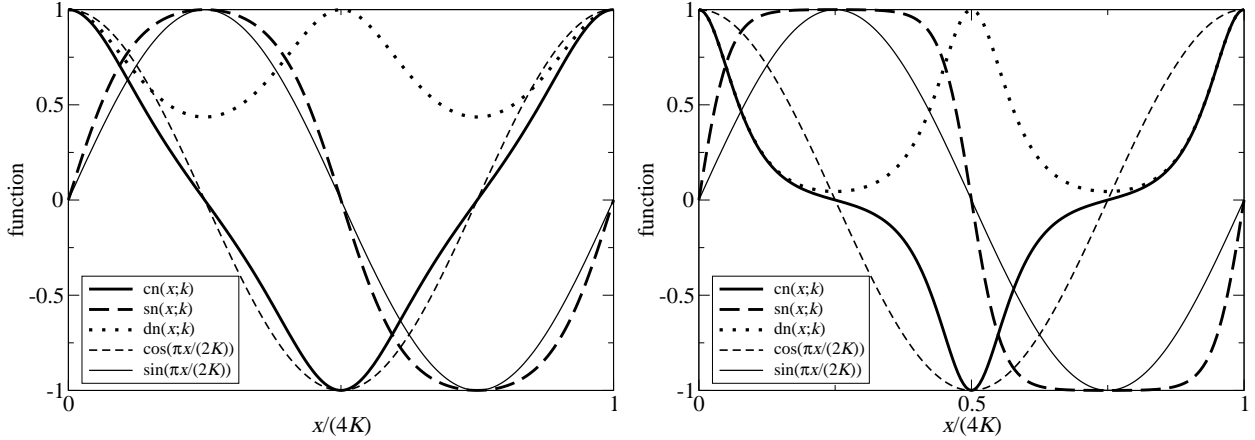

\centerline{\hfill(a) $m=0.81$\hfill\hfill(b) $m=0.998$\hfill}
\centerline{%
\includegraphics[width=0.5\columnwidth]{figure1a}~
\includegraphics[width=0.5\columnwidth]{figure1b}%
}
\caption{Examples of the elliptic functions $\cn$ (solid line), $\sn$
  (bold dashed line), $\dn$ (dotted line) for (a) $m=0.81$ ($K=
  2.28\ldots$, $q= 0.10\ldots$) and (b) $m=0.998$ ($K=4.50\ldots$,
  $q=0.33\ldots$). Also plotted are the cosine (short dashed line) and
  sine (thin short dashed line) with the same period, for comparison.}
\label{cnsndn}
\end{figure}

To give an idea of the behavior of elliptic functions,
figure~\ref{cnsndn} shows a plot of the elliptic functions for two
values of the elliptic parameter, $m=0.81$ and $m=0.998$. It is
evident from the plots that the elliptic functions are periodic
functions of their first argument, and resemble the sine, cosine and
the constant function $1$ unless the value of $m$ is close to one. In
other words, the elliptic parameter $m$ determines how closely the
elliptic functions resemble their trigonometric counterparts.  For
$m=0$, the functions $\cn$, $\sn$ and $\dn$ reduce to $\cos$, $\sin$
and $1$, respectively.  Note that the constant function~$1$ is
reminiscent of the conservation of $\tilde{\omega}_{3}$
(cf.~Eq.~\eqref{symo}) in the case of the symmetric top. Indeed, for
$I_{1}=I_{2}$, Eq.~\eqref{modulus} shows that $m=0$.

Three more numbers can be derived from the elliptic parameter $m$
which play an important role in the properties of elliptic
functions. These are the \emph{quarter-period} $K = F(1|m)$, the
\emph{complementary quarter-period} $K'=F(1|1-m)$ and the \emph{nome}
$q=\exp(-\pi K'/K)$, which is a parameter that appears in various
series expansions of elliptic functions.  In fact, the period of the
elliptic functions $\cn$ and $\sn$ is equal to $4K$, while that of
$\dn$ is $2K$.

Given the solution of the Euler equations in
Eqs.~\eqref{omega2t}--\eqref{omega3t}, the matrix $\mathbf{T}_{1}'$ in
Eq.~\eqref{R1} is completely specified.  On the other hand, to solve
for the time dependence of $\mathbf{T}_{2}'$ in Eq.~\eqref{R2}, the
integral in Eq.~\eqref{psi} must be performed.  Unlike to previous two
cases, the integrand $\Omega$ of $\psi$, as given by
Eq.~\eqref{Omegae}, is not a constant. Despite this complication, the
integral can still be performed explicitly using some properties of
elliptic functions as we will now show. It will require a bit of
complex analysis to integrate $\Omega$ over time, so the reader may
wish to skip this technical part and move on to the answer in
Eq.~\eqref{answer} on page~\pageref{answer} (where the function
$\vartheta_{1}$ is the first Jacobi
function\cite{WhittakerWatson,AbramowitzStegun}).

By definition, an elliptic function is a complex function $f$ (of a
complex variable $t$) that is doubly periodic in the sense that
$f(t+\tau)=f(t+\tau')=f(t)$, and whose singularities in the finite
complex plane consist only of poles.  The doubly periodic function is
completely specified by the values it takes inside the
\emph{fundamental parallelogram} spanned by $\tau$ and $\tau'$ in the
complex plane. Moreover, it may be shown that an elliptic function is
determined by the singularities inside that parallelogram up to an
additive constant\cite{WhittakerWatson,Knopp2}. In fact, for a
elliptic function $f(t)$ having $n$ poles of order one at $t=t_{\rm
pole}^{(j)}$ ($j=1,\ldots,n$) in the fundamental parallelogram with
residues $r_{j}$, one may write (Ref.~\citeonline{WhittakerWatson}, \S
21.5)
\begin{equation}
 f(t) = A_{2}+\sum_{j=1}^{n}  r_{j}
   \frac{d}{dt}
   \log \vartheta_{1}\bigg(\frac{\pi( t-t_{\rm pole}^{(j)})}{\tau}\bigg|m\bigg),
 \label{theta1expr}
\end{equation}
where $A_{2}$ is an additive constant and $\vartheta_{1}(u|m)$ is the first
Jacobi theta function\cite{WhittakerWatson,AbramowitzStegun}.

In order to apply Eq.~\eqref{theta1expr}, the singularity structure of
$\Omega$ must be examined.  A little analysis shows that the function
$\dn$ appearing in $\tilde{\omega}_{3}$ in Eq.~\eqref{omega3t} has
only two poles of order one with opposite residues in its fundamental
parallelogram, and its periods are $\omega_{p}\tau=2K$ and
$\omega_{p}\tau'=4iK'$ (Ref.~\citeonline{AbramowitzStegun}, \S 16.2).
Since any rational function of elliptic functions is again an elliptic
function, the function $\Omega$ in Eq.~\eqref{Omegae} is also an
elliptic function with the same periods. From the form of $\Omega$ in
Eq.~\eqref{Omegae}, one easily sees that any pole $t_{\rm pole}$ of
$\Omega$ in the complex plane is due to a zero of the denominator,
i.e.,
\begin{equation}
  \tilde{\omega}_{3}(t_{\rm pole}) = \pm\frac{L}{I_{3}}.        \label{solveme}
\end{equation}
Note that the poles of $\tilde{\omega}_{3}$ itself cancel in the
numerator and the denominator in Eq.~\eqref{Omegae} leading to a limit
value of $L/I_{3}$, and do not to lead to poles in~$\Omega$.

Using Eq.~\eqref{omega3t}, Eq.~\eqref{solveme} is solved by
\begin{equation}
  \omega_{p}t_{\rm pole} =  \pm\dn^{-1} \left(\pm\frac{L}{I_{3}\tilde{\omega}_{3m}}\bigg|m\right) 
-\varepsilon +  2Kn_{1}+4K'n_{2}i,
\label{pole1}
\end{equation}
Here $n_{1}$ and $n_{2}$ are arbitrary integers and the two $\pm$
signs are independent, thus denoting four possibilities.  Since $\dn$
changes sign when its argument is shifted over a half-period
$2K'i$\cite{AbramowitzStegun}, one $\pm$ sign in Eq.~\eqref{pole1} can
be eliminated by changing the term $+4K'n_{2}i$ to $+2K'n_{2}i$:
\begin{equation*}
  \omega_{p}t_{\rm pole} = \pm\dn^{-1}\left(\frac{L}{I_{3}\tilde{\omega}_{3m}}\bigg|m\right)
  -\varepsilon + 2Kn_{1}+2K'n_{2}i.
\end{equation*}
Using that 
\begin{equation*}
 \dn^{-1}(x|m) = i\left[K'-F(x^{-1}| 1-m)\right],
\end{equation*}
(obtained by combining \S16.3.3, \S16.20.3 and \S17.4.46 in
Ref.~\citeonline{AbramowitzStegun}), one finds
\begin{equation*}
  \omega_{p}t_{\rm pole}=
\pm i\left[K'-F\left(\frac{I_{3}\tilde{\omega}_{3m}}{L}\bigg| 1-m\right)\right]
 -\varepsilon 
+ 2Kn_{1}+2K'n_{2}i .
\end{equation*}
Noting that for negative values of $\tilde{\omega}_{3m}$, one can
write $K'-F(I_{3}\tilde{\omega}_{3m}/L| 1-m)$
$=\sgn(\tilde{\omega}_{3m})K'-F(I_{3}\tilde{\omega}_{3m}/L| 1- m)
+2K'$, we can rewrite this as
\begin{align}
  \omega_{p}t_{\rm pole}&=
\pm i \eta -\varepsilon + 2Kn_{1}+2K'n_{2}i,
\label{pole2}
\end{align}
where we have defined
\begin{equation}
\eta
=\sgn(\tilde{\omega}_{3m})K'-F\left(\frac{I_{3}\tilde{\omega}_{3m}}{L}\bigg|1-m\right)
.
\end{equation}
Note that $-K'\leq\eta\leq K'$.

\begin{figure}[h]
\centerline{\includegraphics[height=0.35\textheight]{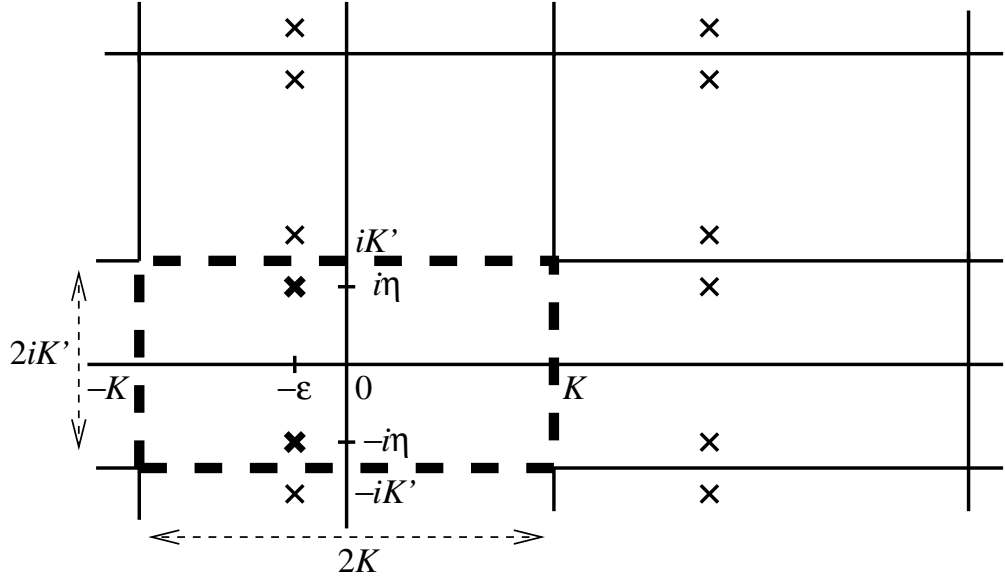}}
\caption{Periodic pole structure in the complex plane of the function
  $\Omega(t)$ as a function of the variable $u=\omega_{p} t$.  The
  fundamental parallelogram is indicated by a bold dashed rectangle
  and the crosses are poles.}\label{poles}
\end{figure}

The periodic structure in Eq.~\eqref{pole2} can be understood as
follows.  The function $\Omega$ depends on~$t$ though $\dn^{2}$, and
$\dn$ has periods $2K$ and $4K'i$.  Note that even though $\dn$
changes sign when its argument is shifted over $2K'i$, this sign
change leaves $\dn^{2}$ and hence $\Omega$ unchanged. From these
considerations, it is evident that the actual periods of $\Omega$ are
$2K/\omega_{p}$ and $2K'i/\omega_{p}$.  Although the size of the
fundamental parallelogram is dictated by the function under
consideration, the choice of its origin is free. Choosing the
fundamental parallelogram to be $([-K-K'i]/\omega_{p},
[-K+K'i]/\omega_{p})$, the two poles in the parallelogram are complex
conjugates $t_{\rm pole}$ and $t_{\rm pole}^{*}$, where
\begin{align}
\omega_{p}t_{\rm pole}&=-\varepsilon+i\eta
.
\label{T}
\end{align}
The pole structure of the function $\Omega(t)$ is illustrated in
Fig.~\ref{poles}.

It is straightforward to show that the residues of $\Omega$ at these
poles are given by $r= \frac{i}{2}$ for the pole at $t_{\rm pole}$ and
$-r= -\frac{i}{2}$ for the pole at $t_{\rm pole}^{*}$, so that, using
Eq.~\eqref{theta1expr} with $\tau=2K/\omega_{p}$, the integrand
$\Omega$ may be written as
\begin{align}
  \Omega &= 
A_{2} + \frac i2\frac{d}{dt}\Big[
 \log\vartheta_{1}\Big(\frac{\pi}{2K}(\omega_{p}t+\varepsilon-i\eta)\Big|m\Big) 
- \log\vartheta_{1}\Big(\frac{\pi}{2K}(\omega_{p}t+\varepsilon+i\eta)\Big|m\Big)
     \Big],
                     \label{thetaexprr}
\end{align}
and further manipulated to obtain a form that can be easily
integrated,
\begin{align}
  \Omega &= 
 A_{2} + \frac i2\frac{d}{dt}\Big[
 \log\vartheta_{1}\Big(\frac{\pi}{2K}(\omega_{p}t+\varepsilon-i\eta)\Big|m\Big)  
- \left\{\log \vartheta_{1}\Big(\frac{\pi}{2K}(\omega_{p}t+\varepsilon-i\eta)\Big|m\Big) \right\}^{*}
     \Big]
\nonumber                   
\\*&=
 A_{2}  -\frac{d}{dt}\im
 \log \vartheta_{1}\Big(\frac{\pi}{2K}(\omega_{p}t+\varepsilon-i\eta)\Big|m\Big) 
\nonumber\\*&
=
 A_{2}  -\frac{d}{dt}\arg \vartheta_{1}\Big(\frac{\pi}{2K}(\omega_{p}t+\varepsilon-i\eta)\Big|m\Big) .
                                         \label{thetaexpr}
\end{align}
The constant $A_{2}$ appearing in Eq.~\eqref{thetaexpr} can be
obtained using the point $t=t_{0}=-\varepsilon/\omega_{p}$, at which
time $\tilde{\omega}_{2}=0$ (cf.~Eq.~\eqref{omega2t}) and
Eq.~\eqref{Omegae} gives
\begin{equation}
  \Omega(t_{0}) =\frac{L}{I_{1}}.
\label{Omegais}
\end{equation}
Examining the right hand side of Eq.~\eqref{thetaexpr}, noting that
\begin{align*}
\frac{d}{dt}\log \vartheta_{1}\Big(\frac{\pi}{2K}(\omega_{p}t+\varepsilon-i\eta)\Big|m\Big) 
&=\frac{\pi\omega_{p}}{2K}\frac{d}{du}
\log \vartheta_{1}\left(u=\frac{\pi}{2K}(\omega_{p}t+\varepsilon-i\eta)
\Big|m\right)
\\&=
\frac{\pi\omega_{p}}{2K}
\frac{\vartheta_{1}'\Big(\frac{\pi}{2K}(\omega_{p}t+\varepsilon-i\eta)\Big|m\Big)}
{\vartheta_{1}\Big(\frac{\pi}{2K}(\omega_{p}t+\varepsilon-i\eta)\Big|m\Big)}
,
\end{align*} 
and using $\vartheta_{1}(-u|m)=-\vartheta_{1}(u|m)$,
Eq.~\eqref{thetaexprr} gives
\begin{equation} 
  \Omega(t_{0}) = A_{2} - \frac{\pi\omega_{p}}{2K}
  \frac
{\vartheta_{1}'\big(\frac{i\pi\eta}{2K}\big|m\big)}
{\vartheta_{1}\big(\frac{i\pi\eta}{2K}\big|m\big)}.
\label{Omegatoo}
\end{equation}
Comparing Eqs.~\eqref{Omegais} and \eqref{Omegatoo}, we see that 
\begin{equation}
  A_{2} = \frac{L}{I_{1}}+  i 
\frac{\pi\omega_{p}}{2K}
  \frac
{\vartheta_{1}'\big(\frac{i\pi\eta}{2K}\big|m\big) }
{\vartheta_{1}\big(\frac{i\pi\eta}{2K}\big|m\big)}.
\label{A2}
\end{equation}
Equation~\eqref{thetaexpr} is now readily integrated to express the
angle $\psi$ in terms of the theta function as:
\begin{align}
  \psi &=\int_{0}^{t}\!dt' \:\: \Omega (t') 
 \nonumber\\*&=
  A_{1}+A_{2}t -
\arg\vartheta_{1}\Big(\frac{\pi}{2K}(\omega_{p}t+\varepsilon-i\eta)\Big|m\Big)  , 
\label{answer}
\end{align}
where
\begin{equation}
A_{1} =
\arg\vartheta_{1}\Big(\frac{\pi(\varepsilon-i\eta)}{2K}\Big|m\Big) ,
\label{A1}
\end{equation}
and $A_{2}$ is given by Eq.~\eqref{A2}.

With this expression for $\psi$, the matrix $\mathbf{P}=$
$\mathbf{T}_{1}'\, \mathbf{T}_{2}'\, \mathbf{T}_{1}{\prime\dagger}(0)$
is now fully specified, with $\mathbf{T}_{1}$ and $\mathbf{T}_{2}$
given in Eqs.~\eqref{R1} and \eqref{R2}, respectively. Unlike the
special cases of spherical and symmetric rotors, no further
simplifications occur in the expressions for these matrices.

\section{Numerical issues}
\label{numerical}

In this section an efficient implementation of computing the exact
solutions presented in the previous section is outlined.  Because the
rotation matrices for spherical and symmetric tops in
Eqs.~\eqref{Pmatrix} and \eqref{Pmatrixsym} are easily implemented
using Rodrigues' formula\cite{Goldstein}, we focus on the case of an
asymmetric rigid rotor. For this case, the attitude matrix was
expressed in terms of complex theta functions, which hinders a
straightforward and efficient numerical implementation. First it will
be shown how all matrix elements in the attitude matrix may expressed
in terms of real quantities and implemented efficiently. At the end of
the section, an algorithm to compute the motion of an asymmetric rigid
body will be presented.

\subsection{Implementation}

As discussed above, the matrix $\mathbf{P}$ can generally be written
as a product of two rotation matrices in the form $\mathbf{T}_{1}\,
\mathbf{T}_{2}$, or three rotations in the form $\mathbf{T}'_{1}\,
\mathbf{T}'_{2}\mathbf{T}_{1}^{\prime\dagger}(0)$. For the
implementation of the free motion of the asymmetric top, we prefer the
latter because it does not require Rodrigues' formula to be applied
twice and the matrix $\mathbf{T}_{2}'$ is sparse which allows an
efficient matrix multiplication. The extra matrix multiplication with
$\mathbf{T}_{1}^{\prime\dagger}(0)$ can be circumvented by using
$\mathbf{B}=\mathbf{T}_{1}^{\prime\dagger}(0)\mathbf{A}(0)$.  The
matrix $\mathbf{T}'_{1}$ is given in Eq.~\eqref{R1} in terms of the
components of the angular momentum in the body frame,
$\tilde{L}_{j}=I_{j}\tilde{\omega}_{j}$. The $\tilde{\omega}_{j}$ are
given by Eqs.~\eqref{omega1t}--\eqref{omega3t}, which involve the
elliptic functions.

Although the occurrence of elliptic functions may seem complicated,
there are standard numerical methods to calculate the elliptic
functions $\sn$, $\cn$ and
$\dn$\cite{AbramowitzStegun,gsl,NumRecipesC} that are very efficient
and which makes them no more problematic to use than standard
transcendental functions such as $\sin$ or $\cos$.  The matrix
$\mathbf{T}_{1}'$ can therefore be computed numerically in a
straightforward way in terms of the standard elliptic functions.

The matrix $\mathbf{T}'_{2}$ is given in Eq.~\eqref{R2} and is a
rotation by an angle $\psi$ around the $z$-axis, with $\psi$ given in
\eqref{answer}. As is evident from Eq.~\eqref{R2}, only $\sin\psi$ and
$\cos\psi$ rather than the angle $\psi$ itself need to be evaluated to
construct $\mathbf{T}_{2}'$.  Using Eq.~\eqref{answer} and the
addition formulas for $\cos$ and $\sin$, these may be expressed as
\begin{align}
 \cos\psi &= \cos(A_{1}+A_{2}t)\cos\arg \vartheta_{1} +\sin(A_{1}+A_{2}t)\sin\arg \vartheta_{1}
 \nonumber\\*&=
 \frac{\cos(A_{1}+A_{2}t)\re \vartheta_{1}+\sin(A_{1}+A_{2}t)\im \vartheta_{1}}
 {\sqrt{(\re \vartheta_{1})^{2}+(\im \vartheta_{1})^{2}}}
\label{cosphi}
\\
 \sin\psi &= 
 \sin(A_{1}+A_{2}t)\cos\arg u-\cos(A_{1}+A_{2}t)\sin\arg u
 =
 \nonumber\\*&=
 \frac{\sin(A_{1}+A_{2}t)\re \vartheta_{1}-\cos(A_{1}+A_{2}t)\im \vartheta_{1}}
 {\sqrt{(\re \vartheta_{1})^{2}+(\im \vartheta_{1})^{2}}}.
\label{sinphi}
\end{align}
where $\vartheta_{1} = \vartheta_{1}\big(\frac{\pi}{2K} (\omega_{p}t +
\varepsilon - i\eta) | m\big)$.  Clearly, to compute these
expressions, the real and imaginary parts of $\vartheta_{1}$ as well as
the real constants $A_{1}$ and $A_{2}$ must be computed.

Noting that the function $\vartheta_{1}$ has the following series
expansion in the nome $q$ (Ref.~\citeonline{AbramowitzStegun}, \S
16.27.1),
\begin{equation}
  \vartheta_{1}(u|m) =
  2q^{1/4}\sum_{n=0}^{\infty}(-1)^{n} q^{n(n+1)}\sin[(2n+1)u],
\label{theta1expansion}
\end{equation}
the real and imaginary parts of $\vartheta_{1}$ in
Eq.~\eqref{theta1expansion} for a complex argument
$u=\frac{\pi}{2K}(\omega_{p}t+\varepsilon-i\eta)$, can be written as
\begin{equation}
\label{reimtheta1}
\begin{split}
  \re \vartheta_{1} &=
2q^{1/4}\sum_{n=0}^{\infty}(-1)^{n} q^{n(n+1)}
\cosh \frac{(2n+1)\pi\eta}{2K}
\sin \frac{(2n+1)\pi(\omega_{p} t+\varepsilon)}{2K}
\\
  \im \vartheta_{1} &= -
2q^{1/4}\sum_{n=0}^{\infty}(-1)^{n} q^{n(n+1)}
\sinh \frac{(2n+1)\pi\eta}{2K}
\cos \frac{(2n+1)\pi(\omega_{p} t+\varepsilon)}{2K}.
\end{split}
\end{equation}
The convergence of these series is extremely rapid due to the
appearance of the $q^{n(n+1)}$. In practice one rarely needs more than
three or four terms to get to machine precision.

Based on the series expansion of $\vartheta_{1}$, the constant $A_{1}$
given in Eq.~\eqref{A1} can be evaluated as follows:
\begin{align}
  A_{1} = n\pi+
  \arctan \frac{\im\vartheta_{1}(\frac{\pi}{2K}(\varepsilon-i\eta)|m)}
{\re\vartheta_{1}(\frac{\pi}{2K}(\varepsilon-i\eta)|m)},
\end{align}
where $n=0$ if $\re\vartheta_{1}>0$, $n=1$ if $\re\vartheta_{1}<0$ and
$\im\vartheta>0$, and $n=-1$ if $\re\vartheta_{1}<0$ and
$\im\vartheta_{1}<0$.

For the constant $A_{2}$, an expansion of the logarithmic derivative
of $\vartheta_{1}$ can be utilized(Ref.~\citeonline{AbramowitzStegun},
\S 16.29.1):
\begin{equation}
\frac{\vartheta_{1}'(u|m)}{\vartheta_{1}(u|m)}
  = \cot u + 4\sum_{n=1}^{\infty}\frac{q^{2n}}{1-q^{2n}}\sin 2nu.
\end{equation}
Using the expression for $A_{2}$ in Eq.~\eqref{A2}, where
$u=i\pi\eta/(2K)$ is purely imaginary and noting that $\cot i u =
-i\coth u=(e^{2u}-1)/(e^{2u} +1)$ and $\sin i u=i\sinh
u=(e^{u}-e^{-u})/2$, one can write
\begin{equation}
i\frac{\vartheta_{1}'(iu|m)}{\vartheta_{1}(iu|m)}
  = \frac{e^{2u}+1}{e^{2u}-1} -2\sum_{n=1}^{\infty}\frac{q^{2n}}{1-q^{2n}}(e^{2nu}-e^{-2nu}).
\label{logderiv}
\end{equation}
Using the series expansion in Eq.~\eqref{logderiv}, $A_{2}$ can be
evaluated from
\begin{equation}
  A_{2} = \frac{L}{I_{1}}+\frac{\pi\omega_{p}}{2K}
\left[ \frac{\xi+1}{\xi-1}-2\sum_{n=1}^{\infty}  \frac{q^{2n}}{1-q^{2n}}(\xi^{n}-\xi^{-n})\right],
\label{xiseries}
\end{equation}
where
\begin{equation}
  \xi=e^{\pi\eta/K}.
\end{equation}
The series in Eq.~\eqref{xiseries} converges if $\xi q^{2}<1$. Because
$-K<\eta<K'$ and $q=\exp(-\pi K'/K)$, one has $\xi q^{2}<q<1$ so this
series converges, and, because $q$ is typically small, usually
quickly.

The above derivation assumed that the Jacobi ordering of
Eq.~\eqref{sure} was satisfied. Although this can always be realized
by choosing which principal axis of the body to call the first, second
or third, as is evident from Eq.~\eqref{sure}, the choice of principal
axes depends on the initial values of the angular velocities.  Often,
instead of choosing the axis depending on the initial conditions, it
is preferable to work with a fixed convention in which the principal
axes are oriented in a particular way with respect to the masses of
the body. In that case, one can adopt the Jacobi ordering convention
by introducing internal variables which differ when necessary from the
physical ones by a rotation. For instance, if $I_{1}>I_{2}>I_{3}$ but
it should be $I_{1}<I_{2}<I_{3}$ according to Eq.~\eqref{sure}, one
can apply the rotation matrix
\begin{equation}
\mathbf{U}^{*} \equiv 
\begin{pmatrix}0&0&1\\0&-1&0\\1&0&0\end{pmatrix},
\label{Ustar}
\end{equation}
which transforms between the internal and physical choices of
principal axes by exchanging the $x$ and $z$ components and reversing
the $y$ component.  If the order of the moments of inertia already
follows the Jacobi ordering, $\mathbf{U}^{*}$ is effectively the
identity matrix.

\subsection{Algorithm for the asymmetric top}
\label{algorithm}

Based on the above, the following algorithm can be set up to calculate
the position of a rigid body at any arbitrary time, given a set of
initial conditions. For efficiency, the algorithm consists of two
steps: an initialization routine, in which some expressions are
pre-calculated, and an evolution routine that calculates
$\tilde{\boldsymbol{\omega}}$ and $\mathbf{A}$ at time $t$.

In the algorithm below, the functions $\sn$, $\cn$, $\dn$ and $F$ are
assumed to be available, but not the $\vartheta_{1}$ function. See
Refs.~\citeonline{gsl,NumRecipesC} for implementations of $\sn$,
$\cn$, $\dn$ and $F$.

The initialization routine takes an initial angular velocity vector in
the body frame $\tilde{\boldsymbol{\omega}}(0)=(\omega_{x0},
\omega_{y0}, \omega_{z0})$ and inertial moments $I_{x}$, $I_{y}$ and
$I_{z}$ (where it is assumed that $I_{y}$ is the middle one), and
pre-computes a few variables as follows (in pseudo-code, in order to
facilitate implementations in different programming languages):

\begin{tt}\parindent 0pt
  Initialization($I_{x}$, $I_{y}$, $I_{z}$, 
  $\tilde{\omega}_{x0}$, $\tilde{\omega}_{y0}$, $\tilde{\omega}_{z0}$, $\mathbf{A}(0)$)
  \begin{tab}
    COMPUTE $L^{2} = I_{x}^{2}\tilde{\omega}_{x0}^{2}+ I_{y}^{2}\tilde{\omega}_{y0}^{2}+ I_{z}^{2}\tilde{\omega}_{z0}^{2}$
    \\
    COMPUTE $2E = I_{x}\tilde{\omega}_{x0}^{2}+ I_{y}\tilde{\omega}_{y0}^{2}+ I_{z}\tilde{\omega}_{z0}^{2}$    \\
    IF ($2E>L^{2}/I_{y}$ AND $I_{x}<I_{z}$) OR ($2E<L^{2}/I_{y}$ AND $I_{x}>I_{z}$) THEN
    \begin{tab}
      SET orderflag\\
      SET $I_{1}=I_{z}$, $I_{2}=I_{y}$ and $I_{3}=I_{x}$\\
      SET $\omega_{10}=\omega_{z0}$, $\omega_{20}= -\omega_{y0}$ and
      $\omega_{30} =  \omega_{x0}$\\
      COMPUTE $L_{\perp} = \sqrt{I_{1}^{2}\omega_{10}^{2}+I_{2}^{2}\omega_{20}^{2}}$\\
      COMPUTE
      $[\mathbf{T}_{1}^{\prime\dagger}(0)\, \mathbf{U}^{*}] =\begin{pmatrix}
      -\frac{L_{\perp}}{L}
      &
      -\frac{I_{2}I_{3}\omega_{20}\omega_{30}}{LL_{\perp}}
      &
            \frac{I_{1}I_{3}\omega_{10}\omega_{30}}{LL_{\perp}}
      \\
      0
      &
      -\frac{I_{1}\omega_{10}}{L_{\perp}}
      &
      -\frac{I_{2}\omega_{20}}{L_{\perp}}
      \\
      \frac{I_{3}\omega_{30}}{L}
      &
      -\frac{I_{2}\omega_{20}}{L}
      &
      \frac{I_{1}\omega_{10}}{L}
      \end{pmatrix}
      $
    \end{tab}
    ELSE
    \begin{tab}
      UNSET orderflag\\
      SET $I_{1}=I_{x}$, $I_{2}=I_{y}$ and $I_{3}=I_{z}$ \\
      SET $\omega_{10}=\omega_{x0}$, $\omega_{20}= \omega_{y0}$ and
      $\omega_{30} =  \omega_{z0}$\\
      COMPUTE $L_{\perp} = \sqrt{I_{1}^{2}\omega_{10}^{2}+I_{2}^{2}\omega_{20}^{2}}$\\
      COMPUTE
      $
      [\mathbf{T}_{1}^{\prime\dagger}(0)\, \mathbf{U}^{*}] =\begin{pmatrix}
      \frac{I_{1}I_{3}\omega_{10}\omega_{30}}{LL_{\perp}}
      &
      \frac{I_{2}I_{3}\omega_{20}\omega_{30}}{LL_{\perp}}
      &
      -\frac{L_{\perp}}{L}
      \\
      -\frac{I_{2}\omega_{20}}{L_{\perp}}
      &
      \frac{I_{1}\omega_{10}}{L_{\perp}}
      &
      0
      \\
      \frac{I_{1}\omega_{10}}{L}
      &
      \frac{I_{2}\omega_{20}}{L}
      &
      \frac{I_{3}\omega_{30}}{L}
      \end{pmatrix}
      $
    \end{tab}
    END IF\\
    COMPUTE $\mathbf{B}=[\mathbf{T}_{1}^{\prime\dagger}(0)\, \mathbf{U}^{*}]\, \mathbf{A}(0)$
    \\
    COMPUTE $\omega_{1m}=
    \sgn(\omega_{10})\sqrt{(L^{2}-2EI_{3})/(I_{1}(I_{1}-I_{3}))}$  \\
    COMPUTE $\omega_{2m}=-\sgn(\omega_{10})\sqrt{(L^{2}-2EI_{3})/(I_{2}(I_{2}-I_{3}))}$
    \\
    COMPUTE $\omega_{3m}=\sgn(\omega_{30})\sqrt{(L^{2}-2EI_{1})/(I_{3}(I_{3}-I_{1}))}$\\
    COMPUTE $\omega_{p} =\sgn(I_{2}-I_{3})\sgn(\omega_{30})\sqrt{(L^{2}-2EI_{1})(I_{3}-I_{2})/(I_{1}I_{2}I_{3})}$\\
    COMPUTE $m = (L^{2}-2EI_{3})(I_{1}-I_{2})/((L^{2}-2EI_{1})(I_{3}-I_{2}))$\\
    COMPUTE $\varepsilon =  F(\omega_{20}/\omega_{2m} | m)$\\
    COMPUTE $K =  F(1 | m)$\\
    COMPUTE $K'=  F(1 | 1-m)$\\
    COMPUTE $q =  \exp(-\pi K'/K)$\\
    COMPUTE $\eta = \sgn(\omega_{30}) K'-F(I_{3}\omega_{3m}/L,1-m)$\\
    COMPUTE $\xi  =\exp(\pi\eta/K)$\\
    SET $\displaystyle A_{2} = L/I_{1}+\pi\omega_{p}(\xi+1)/(2K(\xi-1))$\\
    SET $n=1$\\
    REPEAT
    \begin{tab}
      COMPUTE $\displaystyle\delta A_{2} = -(\pi\omega_{p}/K)
      (q^{2n}/(1-q^{2n}))(\xi^{n}-\xi^{-n})$\\
      INCREMENT $A_{2}$ by $\delta A_{2}$\\
      INCREMENT $n$ by $1$
    \end{tab}
    UNTIL $\delta A_{2} >$ machine precision\\
    COMPUTE $NT = \log(\text{machine precision})/\log q$\\
    SET $r_{0} = 0$ and  $i_{0} = 0$\\
    FOR $n=0$ TO $NT$ DO
    \begin{tab}
      COMPUTE $c_{r}[n]=(-1)^{n}2q^{n(n+1)+1/4}
      \cosh\frac{(2n+1)\pi\eta}{2K}$\\
      COMPUTE $c_{i}[n]=(-1)^{n+1}2q^{n(n+1)+1/4}
      \sinh\frac{(2n+1)\pi\eta}{2K}$\\
      INCREMENT $r_{0}$ by $c_{r}[n]\sin\frac{(2n+1)\pi\varepsilon}{2K}$ \\
      INCREMENT $i_{0}$ by $c_{i}[n]\cos\frac{(2n+1)\pi\varepsilon}{2K}$ 
    \end{tab}
    END FOR\\
    IF $r_{0}>0$ THEN $k =0$ ELSE $k = \sgn(i_{0})$\\
    COMPUTE $A_{1}=\arctan(i_{0}/r_{0})+k\pi$\\
    STORE orderflag, $I_{1}$, $I_{2}$, $I_{3}$, $\omega_{1m}, \omega_{2m}, \omega_{3m}, m,
    \omega_{p}, \varepsilon, A_{1}, A_{2}, NT, c_{r}[\,], c_{i}[\,]$ and $\mathbf{B}$
  \end{tab}
  END Initialization
\end{tt}

\noindent
The evolution routine can use the pre-computed expressions in the
following way:

\begin{tt}\parindent 0pt
Evolution($t$)
\begin{tab}
  COMPUTE $\tilde \omega_{1} = \omega_{1m}\cn(\omega_{p} t+\varepsilon|m)$\\
  COMPUTE $\tilde \omega_{2} = \omega_{2m}\sn(\omega_{p} t+\varepsilon|m)$\\
  COMPUTE $\tilde \omega_{3} = \omega_{3m}\dn(\omega_{p} t+\varepsilon|m)$\\
  SET $\re\vartheta_{1}=0$ and $\im\vartheta_{1}=0$   \\
  FOR $n=0$ TO $NT$ DO
  \begin{tab}
    INCREMENT $\re\vartheta_{1}$ by  $c_{r}[n]\sin((2n+1)\pi (\omega_{p} t+\varepsilon)/(2K))$\\
    INCREMENT $\im\vartheta_{1}$ by  $c_{i}[n]\cos((2n+1)\pi (\omega_{p} t+\varepsilon)/(2K))$
  \end{tab}
  END FOR\\
  COMPUTE $C=\cos(A_{1}+A_{2} t)$, $S=\sin(A_{1}+A_{2} t)$\\
  COMPUTE $\cos\psi  = (C\, \re\vartheta_{1}+S\,
    \im\vartheta_{1})/\sqrt{\re\vartheta_{1}^{2}+\im\vartheta_{1}^{2}}$
  \\
  COMPUTE $\sin\psi  = (S\, \re\vartheta_{1}-C\, \im\vartheta_{1})/\sqrt{\re\vartheta_{1}^{2}+\im\vartheta_{1}^{2}}$\\
  COMPUTE $L_{\perp} = \sqrt{I_{1}^{2}\omega_{1}^{2}+I_{2}^{2}\omega_{2}^{2}}$\\
  IF orderflag IS SET THEN
  \begin{tab}
    COMPUTE $[\mathbf{U}^{*}\, \mathbf{T}'_{1}] =\begin{pmatrix}
    -\frac{L_{\perp}}{L}
    &
    0
    &
    \frac{I_{3}\omega_{3}}{L}
    \\
    -\frac{I_{2}I_{3}\tilde{\omega}_{2}\tilde{\omega}_{3}}{LL_{\perp}}
    &
    -\frac{I_{1}\tilde{\omega}_{1}}{L_{\perp}}
    &
    -\frac{I_{2}\tilde{\omega}_{2}}{L}
    \\
    \frac{I_{1}I_{3}\tilde{\omega}_{1}\tilde{\omega}_{3}}{LL_{\perp}}
    &
    -\frac{I_{2}\tilde{\omega}_{2}}{L_{\perp}}
    &
    \frac{I_{1}\tilde{\omega}_{1}}{L}
    \end{pmatrix}
    $\\
    SWAP $\tilde\omega_{1}$ and $\tilde\omega_{3}$\\
    CHANGE SIGN of $\tilde\omega_{2}$
  \end{tab}
  ELSE
  \begin{tab}
    COMPUTE $[\mathbf{U}^{*}\, \mathbf{T}'_{1}] =\begin{pmatrix}
    \frac{I_{1}I_{3}\tilde{\omega}_{1}\tilde{\omega}_{3}}{LL_{\perp}}
    &
    -\frac{I_{2}\tilde{\omega}_{2}}{L_{\perp}}
    &
    \frac{I_{1}\tilde{\omega}_{1}}{L}
    \\
    \frac{I_{2}I_{3}\tilde{\omega}_{2}\tilde{\omega}_{3}}{LL_{\perp}}
    &
    \frac{I_{1}\tilde{\omega}_{1}}{L_{\perp}}
    &
    \frac{I_{2}\tilde{\omega}_{2}}{L}
    \\
    -\frac{L_{\perp}}{L}
    &
    0
    &
    \frac{I_{3}\omega_{3}}{L}
    \end{pmatrix}$
  \end{tab}
  END IF\\
  COMPUTE $\mathbf{A} = [\mathbf{U}^{*}\, \mathbf{T}'_{1}]\begin{pmatrix}
    \cos\psi & \sin\psi &0
    \\
    -\sin\psi&\cos\psi &0
    \\
    0&0&1
    \end{pmatrix}\mathbf{B}$
  \\
  RETURN $\tilde\omega_{1}$, $\tilde\omega_{2}$, $\tilde\omega_{3}$ and
  $\mathbf{A}$
\end{tab}
END Evolution 
\end{tt}\\
Some remarks about this pseudo-code:
\begin{itemize}

\item The {\tt orderflag} indicates whether the Jacobi ordering
  convention Eq.~\eqref{sure} is satisfied. If not, $\mathbf{U}^{*}$
  in Eq.~\eqref{Ustar} is used. If Eq.~\eqref{sure} is satisfied,
  $\mathbf{U}^{*}$ is set equal to the identity matrix.

\item With these definitions, $\mathbf{P}$ is replaced by
  $\mathbf{U}^{*} \mathbf{P}\, \mathbf{U}^{*}$. This is accomplished
  simply by $\mathbf{T}'_{1}\to \mathbf{U}^{*}\, \mathbf{T}'_{1}$.

\item As a consequence, $\mathbf{U}^{*}$ in Eq.~\eqref{Ustar} is only
  implicitly used in the combination $[\mathbf{U}^{*}\,
  \mathbf{T}'_{1}]$

\item The initial value of $\mathbf{A}(0)$ occurs as $\mathbf{A}(t) =
  \mathbf{U}^{*}\, \mathbf{T}'_{1} \, \mathbf{T}'_{2} \,
  \mathbf{T}_{1}^{\prime\dagger}(0)\, \mathbf{U}^{*} \,
  \mathbf{A}(0)$, so in the initialization routine, we only need to
  store the combination
  $\mathbf{B}=\mathbf{T}_{1}^{\prime\dagger}(0)\, \mathbf{U}^{*} \,
  \mathbf{A}(0)$.

\item The {\tt machine precision} depends on the floating point
  precision used in the calculation; for 64 bit double precision, this
  is of the order of $10^{-15}$--$10^{-16}$.

\item For clarity of the algorithm, matrix products have not been
  explicitly written out, and efficiency improvements such as
  computing intermediate expressions and using a recursive evaluation
  for the $\sin$'s and $\cos$'s have not been shown here.

\item An implementation of this code in C, which includes these
  improvements, can be found on the internet at
  \texttt{http://www.chem.utoronto.ca/staff/JMS/rigidrotor.html} and
  in the appendix.
\end{itemize}

\section{Example}
\label{example}

\begin{figure}[t]
\centerline{\hfill(a)\hfill\hfill(b)\hfill}
\centerline{\includegraphics[width=0.45\textwidth]{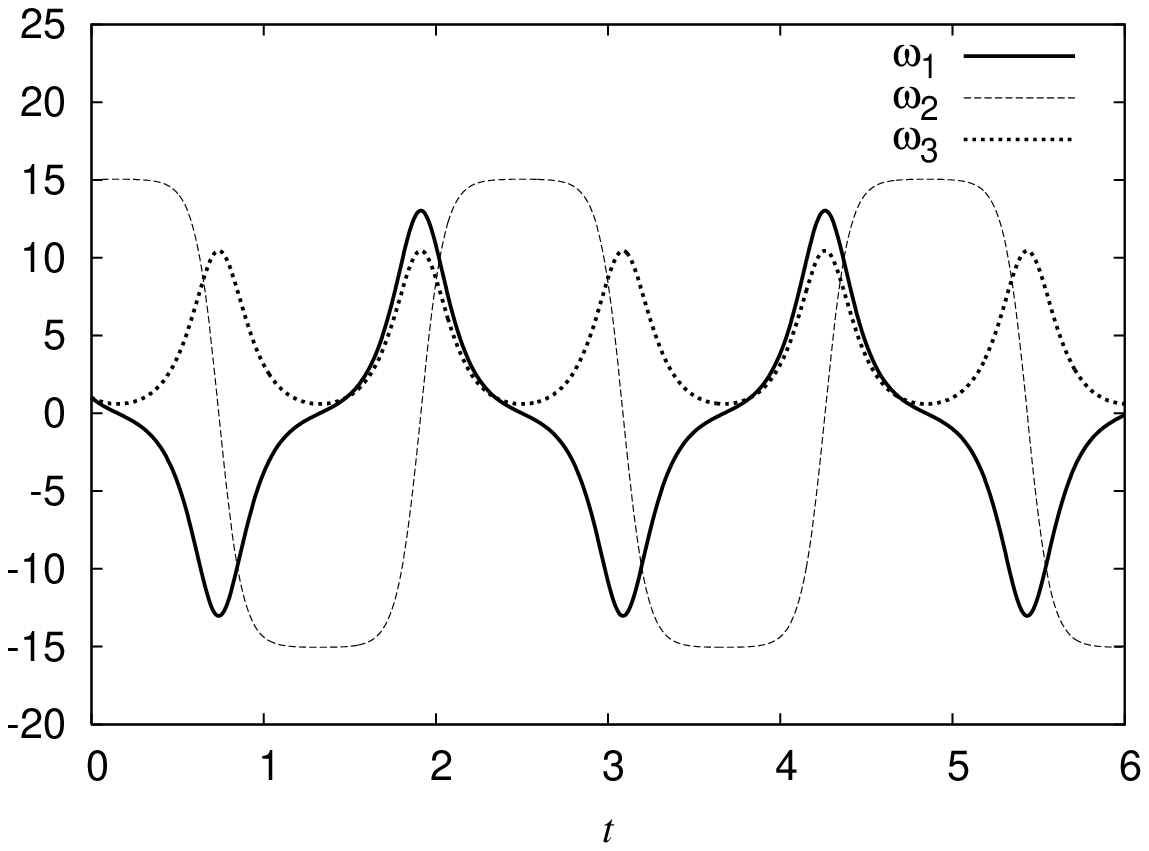}%
\includegraphics[width=0.45\textwidth]{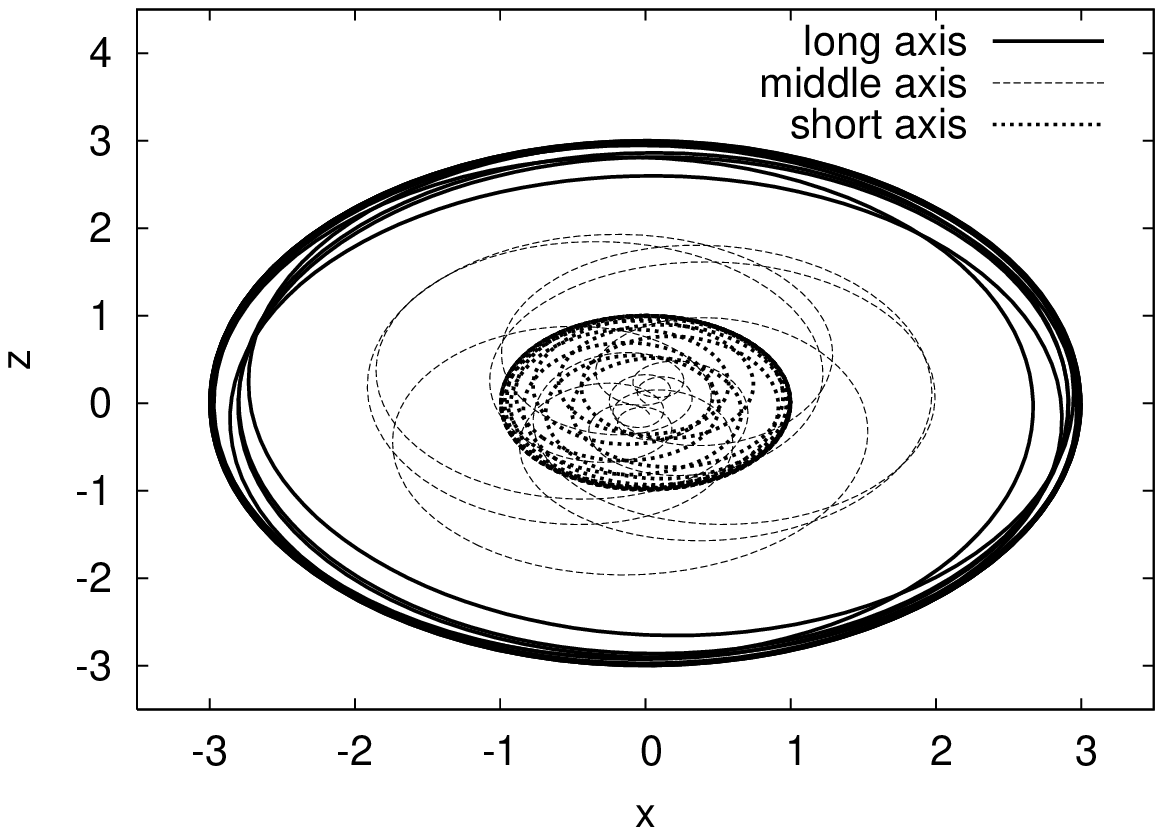}}
\caption{Example of the exact solution. (a) $\tilde\omega(t)$,
(b) position of the `axes'. For details see text.}\label{figure3}
\end{figure}

As an example, consider an object composed of six point masses which
are arranged, in the body frame, at the points $(a,0,0)$, $(-a,0,0)$,
$(0,b,0)$, $(0,-b,0)$, $(0,0,c)$ and $(0,0,-c)$. All points are
assumed to have unit mass, so that the inertial moments are given by
$I_{1}=2(b^{2}+c^{2})$, $I_{2}=2(a^{2}+c^{2})$ and
$I_{3}=2(a^{2}+b^{2})$. Choosing $a>b>c$ ensures $I_{1}<I_{2}<I_{3}$.

In particular we will consider $a=3$, $b=2$ and $c=1$, yielding
$I_{1}=10$, $I_{2}=20$ and $I_{3}=26$. As initial conditions we will
take $\mathbf{A}(0)=\mathbf{1}$ and $\boldsymbol{\omega}(0)=(1,15,1)$.
Because of the large value of the $y$-component of the angular
velocity, one may expect motion to consist primarily of rotation
around the $y$-axis. In figure~\ref{figure3}a the components of the
angular velocity in the body frame have been plotted as a function of
time, while in figure~\ref{figure3}b the projection on the $x-z$ plane
of the point masses that in the body frame are located at $(a,0,0)$
(the `long axis'), $(0,b,0)$ (the `middle' axis') and $(0,0,c)$ (the
`short axis') are plotted. It is clear that the rotational motion does
not consist of small perturbations to a rotation around the
$y$-axis. We stress once more that, up to machine precision, the
results in figure~\ref{figure3} are exact.

\section{Discussion}
\label{discussion}

In this paper, the general solution of the rotational motion of a
rigid body in the absence of external torques and forces was derived.
Explicit expressions for the angular velocities and the attitude
matrix were obtained in terms of real quantities to facilitate
numerical evaluation.  Note that even though the solution of
rotational motion for bodies without a simple mass distribution
contains generalizations of the familiar sine and cosine functions,
the motion typically appears quite complex and notably different from
that of a spherical top.

The general solution of the equations governing rigid body dynamics in
the absence of forces and torques presented here is potentially useful
in several important applications.  The primary advantage of having in
hand analytical solutions of the equations of motion of a system lies
in the fact that all relevant properties of a soluble system can be
determined at arbitrarily many and arbitrarily distant moments in
time.  Applications in which knowing exactly the position and
orientation of a body at specific moments in time is paramount may
benefit from the results presented here.  Such applications are
abundant in a wide variety of contexts.  For example, in astrophysics,
many objects such as space crafts, asteroids, certain planets and
moons, behave on short time scales as rigid bodies. These bodies are
not free since they feel typically weak gravitational fields. However,
if their dimensions are small enough compared to the gradients of the
gravitational field, gravitational forces effectively influence motion
only of the center of mass, while the rotational motion is that of a
free rotor described here.

Another obvious application of the solution detailed in this paper is
as a diagnostic tool for numerical integration techniques designed for
rotating bodies with external torques. Such techniques are of
considerable interest, but to establish their accuracy, one needs to
be able to compare results of approximate integration schemes with
exact results.  To date most comparisons are carried out for free
systems with a high degree of symmetry and simple rotational
motion\cite{rotation}.  Given the relative complexity of motion in the
asymmetric case compared with that of a spherical rotor, such
comparisons do not appear to be very stringent.

The exact solution is also of practical use in symplectic integrators
for use in continuous molecular dynamics\cite{newintegrator}. There
are already various symplectic integrators using the exact solution of
some part of the dynamics\cite{CelledoniSafstrom,appendixC,exactpart},
which generally seems to improve stability and accuracy over simple
splitting
methods\cite{rotation0,rotation,rotation2,mclachlanquispel}. However,
these do not use the exact solution of the attitude matrix. The
integrator of Celledoni and S\"afstr\"om, for example, uses the exact
solution of the Euler equations but uses an approximate expression for
the attitude matrix\cite{CelledoniSafstrom}. Using the exact solution
of the attitude matrix further improves the stability and accuracy of
this integrator\cite{newintegrator}.

Perhaps the most direct application of the implementation of the exact
solution is in simulating complex rigid molecular systems using
discontinuous molecular dynamics methods.  In this approach, various
components of the system interact via discontinuous potentials,
leading to impulsive forces and torques that act on molecules at
specific moments in time\cite{AlderWainwright57,Chapelaetal84,
AllenTildesley,FrenkelMaguire83,Rapaport}.  As a result, the motion of
all bodies in the system is free between impulsive events that alter
the trajectory of the body via discontinuous jumps in the momenta or
angular velocities at discrete ``collision'' times.  In order to
determine the time at which molecules in the simulation interact, the
exact location and orientation of all bodies in the system must be
computable at arbitrary times.  If the configurations of the system
are computed through numerical integration (using e.g. one of the
integrators in
Refs.~\citeonline{rotation0,rotation,rotation2,CelledoniSafstrom}),
such simulations would become inefficient. For this reason, to date,
most simulations of rigid bodies interacting via discontinuous
potentials have been restricted to systems in which rotational motion
is governed by the equations of a spherical rotor (see Refs.\
\citeonline{Chapelaetal84}).  Armed with the results of this paper,
the technique of discontinuous molecular dynamics can now be applied
to any rigid model---symmetric or asymmetric---with discontinuous
interactions of step-potential form.  Examples of such studies can be
found in Refs.\ \citeonline{DMD2}.

\section*{Acknowledgments}

The authors would like to thank Prof.\ Sheldon Opps, Dr.\ Lisandro
Hern\'andez de la Pe\~na and Prof.\ Stuart Whittington for useful
discussions. This work was supported by a grant from the National
Sciences and Engineering Research Council of Canada.

\appendix
\section*{Appendix: Implementation in C}

\begin{verbatim}
#define MACHPREC 3.E-16
#define MY_PI_2 1.57079632679489661923 
/* The following macro performs an in-situ matrix multiplication: */
#define RIGHTMULTMATRIX(M1, M2, x, y)\
  x = M1[0][0]; y = M1[0][1];\
  M1[0][0] *= M2[0][0]; M1[0][0] += y*M2[1][0]+M1[0][2]*M2[2][0];\
  M1[0][1] *= M2[1][1]; M1[0][1] += x*M2[0][1]+M1[0][2]*M2[2][1];\
  M2[2][2]; M1[0][2] += x*M2[0][2]+y*M2[1][2];\
  x = M1[1][0]; y = M1[1][1];\
  M1[1][0] *= M2[0][0]; M1[1][0] += y*M2[1][0]+M1[1][2]*M2[2][0];\
  M1[1][1] *= M2[1][1]; M1[1][1] += r*M2[0][1]+M1[1][2]*M2[2][1];\
  M1[1][2] *= M2[2][2]; M1[1][2] += r*M2[0][2]+y*M2[1][2];\
  x = M1[2][0]; y = M1[2][1];\
  M1[2][0] *= M2[0][0]; M1[2][0] += y*M2[1][0]+M1[2][2]*M2[2][0];\
  M1[2][1] *= M2[1][1]; M1[2][1] += x*M2[0][1]+M1[2][2]*M2[2][1];\
  M1[2][2] *= M2[2][2]; M1[2][2] += x*M2[0][2]+y*M2[1][2];

/* A number of precomputed expressions are taken together in the
   following structure: */
typedef struct {
 int orderFlag, NT;
 double I1, I2, I3, omega1m, omega2m, omega3m, omegap, freq, epsilon, 
        A1, A2, L, m, *cr, *ci, B[3][3];
} Top;

/* The following function performs some pre-calculations. Input: Ix,  *
 * Iy and Iz are the three principal inertial moments (it is assumed  *
 * that Ix<Iy<Iz or Ix>Iy>Iz), omegax, omegay and omegaz are the      *
 * angular velocities in the principal axis (i.e. body) frame, and A  *
 * is the initial attitude matrix. Returns a Top structure containing *
 * all necessary precomputed parameters to efficiently calculate the  *
 * omega's and A's at later times (see the Evolution function below): */
Top Initialization(double Ix, double Iy, double Iz, 
                   double omegax, double omegay, double omegaz, double A[3][3])
{
 double
  a = Ix*omegax,                       /* L1, angular momentum in x-direction */
  b = Iy*omegay,                       /* L2, angular momentum in y-direction */
  c = Iz*omegaz,                       /* L3, angular momentum in z-direction */
  d = a*a+b*b+c*c,                   /* L.L, norm squared of angular momentum */
  e = a*omegax+b*omegay+c*omegaz,                                      /* 2 E */
  f, omega1, omega2, omega3, Kp, q, r, i, s, g, h;
 
 int n;
 Top R;                               /* will contain the precomputed numbers */
 R.L = sqrt(d);                                                        /* |L| */
 if ( (e>d/Iy && Ix<Iz) || (e<d/Iy && Ix>Iz) ) { 
  /* Check if Jacobi-ordering is obeyed: */
  R.orderFlag = 1;/* Jacobi ordering ensured by swapping the order of x and z */
  R.I1 = Iz;                      /* directions and reversing the y direction */
  R.I2 = Iy;
  R.I3 = Ix;
  omega1 = omegaz;
  omega2 = -omegay;
  omega3 = omegax;
  f = hypot(b, c);                                                  
  /* Fill the matrix R.B with transpose of T1(0), using ordering: */
  R.B[0][0] = -f/R.L; R.B[0][1] = b*a/R.L/f; R.B[0][2] = a*c/R.L/f;
  R.B[1][0] = 0;      R.B[1][1] = -c/f;      R.B[1][2] = b/f;
  R.B[2][0] = a/R.L;  R.B[2][1] = b/R.L;     R.B[2][2] = c/R.L;
 } else {
   R.orderFlag = 0;                                /* Jacobi ordering correct */
   R.I1 = Ix;
   R.I2 = Iy;
   R.I3 = Iz;
   omega1 = omegax;
   omega2 = omegay;
   omega3 = omegaz;
   f = hypot(a, b);
   /* Fill the matrix  R.B with the transpose of T1(0): */
   R.B[0][0] = a*c/R.L/f; R.B[0][1] = b*c/R.L/f; R.B[0][2] = -f/R.L;
   R.B[1][0] = -b/f;      R.B[1][1] = a/f;       R.B[1][2] = 0;
   R.B[2][0] = a/R.L;     R.B[2][1] = b/R.L;     R.B[2][2] = c/R.L;
 }
 RIGHTMULTMATRIX(R.B, A, r, i);                               /*  calculate B */
 a = d-e*R.I3;                                 /* compute four subexpressions */
 b = d-e*R.I1;
 c = R.I1-R.I3; 
 d = R.I2-R.I3;
 R.omega1m = copysign(sqrt(a/R.I1/c), omega1); 
 R.omega2m = -copysign(sqrt(a/R.I2/d), omega1);
 R.omega3m = copysign(sqrt(-b/R.I3/c), omega3);
 R.omegap = d*copysign(sqrt(b/(-d)/R.I1/R.I2/R.I3), omega3);    /* prec. freq */
 R.m = a*(R.I2-R.I1)/(b*d);                            /* ellipic parameter m */
 R.epsilon = F(omega2/R.omega2m, R.m);                       /* initial phase */
 R.freq = MY_PI_2/F(1.0, R.m);            /* frequency relative to precession */
 Kp = F(1.0, 1.0-R.m);                    /* K', complementary quarter period */
 q = exp(-2.0*R.freq*Kp);                                    /* elliptic nome */
 e = exp(R.freq*(copysign(Kp, omega3)-F(R.I3*R.omega3m/R.L, 1.0-R.m)));
\end{verbatim}
\verb+                                             /*  = +
$\exp\left[\frac{\pi(\sgn(\omega_3)K'-F(I_3\omega_{3m}/L|m)}{2K}\right]$ \verb+*/+
\begin{verbatim}
 f = e*e;                                            /* this f is equal to xi */
 R.A2 = R.L/R.I1+R.freq*R.omegap*(f+1)/(f-1);      /* first term in A2 series */
 a = 1.0;                                   /* a will be the 2n-th power of q */
 b = 1.0;                                    /* b will be the nth power of xi */
 n = 1;
 do {
  a *= q*q;                                    /* update a and  b recursively */
  b *= f;
  c = -2.0*R.freq*R.omegap*a/(1-a)*(b-1/b);            /* the next term in A2 */
  R.A2 += c;                                      /* add a term of the series */
  n++;
 } while (fabs(c/R.A2)>MACHPREC && n<10000);/* stop if converged or n too big */
 /* determine upper bound on number of terms needed in theta function series: */
 R.NT = (int)(log(MACHPREC)/log(q)+0.5);
 R.cr = (double *)malloc(sizeof(double)*(R.NT+1));         /* allocate memory */
 R.ci = (double *)malloc(sizeof(double)*(R.NT+1)); 
 a = 1.0;                                   /* a will be the 2n-th power of q */
 b = 1.0;                                      /* b will be (-1)^n q^{n(n+1)} */
 R.cr[0] =  (e+1/e);      /* zeroth term in the series for real and imag part */
 R.ci[0] = -(e-1/e); 
 s = sin(R.freq*R.epsilon);           /*  s = sin((2n+1)x),  c = cos((2n+1)x) */
 c = cos(R.freq*R.epsilon); 
 g = 2.0*c*s;                                                      /* sin(2x) */
 h = 2.0*c*c-1.0;                                                  /* cos(2x) */
 r = R.cr[0]*s;               /* real part of the theta function, zeroth term */
 i = R.ci[0]*c;          /* imaginary part of the theta function, zeroth term */
 for (n = 1; n <=  R.NT; n++) {
  e *= f;                                                 /* e  =  xi^{n+1/2} */
  a *= q*q;                                     /* update a and b recursively */
  b *= -a;
  R.cr[n] = b*(e+1/e);            /* compute next coefficient of theta series */
  R.ci[n] = -b*(e-1/e); 
  d = s;                                   /* compute sin and cos recursively */
  s = h*s+g*c; 
  c = h*c-g*d; 
  r += R.cr[n]*s;                                           /* add next terms */
  i += R.ci[n]*c; 
  if ( (fabs(R.cr[n]) < MACHPREC) && (fabs(R.ci[n]) < MACHPREC) ) 
   R.NT = n-1;                                     /* if converged, adjust NT */
 }
 R.A1 = atan2(i, r);                                     /* compute arg(r, i) */
 return R;                             /* done, return all precomputed values */
}


/* The following function calculates the angular velocities and attitude     * 
 * matrix at a time t. Input: R is a pointer to a Top structure containing   * 
 * all necessary precomputed parameters to efficiently calculate the omega's *
 * and A's, and t is a time. R should be generated by Initialize function.   *
 * Output: *omegax, *omegay and *omegaz are filled with the angular          * 
 * velocities in the principal axis (i.e. body) frame at time t and A is     *
 * filled with the attitude matrix at time t.                                */
void Evolution(Top *R, double t, double *omegax, double *omegay, double *omegaz, 
               double A[3][3]) 
{
 int n; 
 double omega1, omega2, omega3, r, i, u, s, c, g, h, f;

 u = R->omegap*t+R->epsilon;        /* compute argument of elliptic functions */
 SNCNDN(u, R->m, &omega2, &omega1, &omega3);            /* compute sn, cn, dn */
 omega1 *= R->omega1m;                         /* multiply by the amplitudes  */
 omega2 *= R->omega2m;                         /* of the respective ang. vel. */
 omega3 *= R->omega3m;
 if (R->orderFlag == 1) {                        /* check for Jacobi ordering */
  *omegax = omega3;              /* if adjusted, invert x and z and reverse y */
  *omegay = -omega2; 
  *omegaz = omega1;
  omega1 *= R->I1; 
  omega2 *= R->I2; 
  omega3 *= R->I3;
  f = hypot(omega1, omega2);  
  /* compute T1(t), taking into account the ordering:  */
  A[0][0] = -f/R->L;               A[0][1] = 0;         A[0][2] = omega3/R->L;
  A[1][0] = -omega2*omega3/R->L/f; A[1][1] = -omega1/f; A[1][2] = -omega2/R->L;
  A[2][0] = omega1*omega3/R->L/f;  A[2][1] = -omega2/f; A[2][2] = omega1/R->L;
 } else {
  *omegax = omega1;                                /* no adjustment necessary */
  *omegay = omega2; 
  *omegaz = omega3;
  omega1 *= R->I1; 
  omega2 *= R->I2; 
  omega3 *= R->I3;
  f = hypot(omega1, omega2);     
  /* compute T1(t): */
  A[0][0] = omega1*omega3/R->L/f; A[0][1] = -omega2/f; A[0][2] = omega1/R->L;
  A[1][0] = omega2*omega3/R->L/f; A[1][1] = omega1/f;  A[1][2] = omega2/R->L;
  A[2][0] = -f/R->L;              A[2][1] = 0;         A[2][2] = omega3/R->L;
 }
 s = sin(R->freq*u);    /* s = sin((2n+1)x), c = cos((2n+1)x) with x = freq*u */
 c = cos(R->freq*u);
 g = 2.0*c*s;                    /* g = sin 2x, h = cos 2x, used in recursion */
 h = 2.0*c*c-1.0;
 r = R->cr[0]*s;              /* zeroth term in series for the theta function */
 i = R->ci[0]*c;                
 for (n = 1; n <= R->NT; n++) {                             /* compute series */
  u = s;
  s = h*s+g*c;           /* computes sin((2n+1)x) and cos((2n+1)x recursively */
  c = h*c-g*u;
  r += R->cr[n]*s;                  /* next term in series for theta function */ 
  i += R->ci[n]*c;
 }
  s = sin(R->A1+R->A2*t);                               /* s = sin(A_1+A_2 t) */
  c = cos(R->A1+R->A2*t);                               /* c = cos(A_1+A_2 t) */
  u = s;       /* use addition formula to compute s = sin psi and c = cos psi */
  s = s*r-c*i;                             /* where psi = A_1+A_2 t+arg(r, i) */
  c = c*r+u*i;
  u = hypot(r, i);
  s /= u;
  c /= u;
  for (n = 0; n<3; n++) {                  /* perform multiplication with T2' */
   u = A[n][0];     
   A[n][0] = A[n][0]*c-A[n][1]*s;
   A[n][1] = A[n][1]*c+u*s;
  }
  RIGHTMULTMATRIX(A, R->B, r, i);         /* gives the  final attitude matrix */
}		
\end{verbatim}

\noindent
Let us briefly mention a few accessible implementations of elliptic functions and
elliptic integrals:
\begin{itemize}
  \item Using the GNU Scientific Library\cite{gsl}:
\begin{verbatim}
double F(double x, double m) 
{ return x*gsl_sf_ellint_RF(1.0-x*x, 1.0-m*x*x, 1.0, GSL_PREC_DOUBLE); }
void SNCNDN(double x, double m, double *s, double *c, double *d) 
{ (void)gsl_sf_elljac_e(x, m, s, c, d); }
\end{verbatim}

  \item Using Numerical Recipes\cite{NumRecipesC}:
\begin{verbatim}
double F(double x, double m) 
{ return x*rf(1.0-x*x, 1.0-m*x*x, 1.0); }
void SNCNDN(double x, double m, double *s, double *c, double *d) 
{ sncndn(x, 1.0-m, s, c, d); }
\end{verbatim}
To truly work in double precision, compared to
Ref.~\citeonline{NumRecipesC}, each \texttt{float} should be replaced by
\texttt{double} and the following constants in the source
code would have to be changed: 
\begin{verbatim}
#define ERRTOL 0.0025
#define CA 1e-8 
\end{verbatim}
in rf and sncndn, respectively.

\end{itemize}

\end{document}